\documentclass[a4paper,11pt]{article}
\usepackage[labelfont=bf,labelsep=period]{caption}
\usepackage[nodayofweek]{datetime}
\usepackage{enumitem}
\usepackage{float}
\usepackage[margin=2.5cm]{geometry}
\usepackage{graphicx}
\usepackage{hyperref}
\usepackage[utf8]{inputenc}
\usepackage{numbertabbing}
\usepackage{times}
\usepackage{url}
\usepackage{xcolor}
\usepackage{xspace}
\usepackage{comment}
\usepackage{subcaption}

\usepackage{amsmath} %%% better math %%%
\allowdisplaybreaks[2]          % but allow some broken displays
\usepackage{amssymb} %%% symbols %%%
\usepackage{amsthm} %%% Theorem environments with `amsthm' %%%
\newtheoremstyle{plain-boldhead}% name
  {\topsep}%      Space above
  {\topsep}%      Space below
  {\itshape}%     Body font
  {}%         Indent amount (empty = no indent, \parindent = para indent)
  {\bfseries}% Thm head font
  {.}%        Punctuation after thm head
  { }%     Space after thm head: " " = normal space; \newline = linebreak
  {\thmname{#1}\thmnumber{ #2}\thmnote{ (\bfseries #3)}}%    Thm head spec
\newtheoremstyle{definition-boldhead}% name
  {\topsep}%      Space above
  {\topsep}%      Space below
  {\normalfont}% Body font
  {}%         Indent amount (empty = no indent, \parindent = para indent)
  {\bfseries}% Thm head font
  {.}%        Punctuation after thm head
  { }%     Space after thm head: " " = normal space; \newline = linebreak
  {\thmname{#1}\thmnumber{ #2}\thmnote{ (\bfseries #3)}}%    Thm head spec
\theoremstyle{plain-boldhead}
\newtheorem{theorem}{Theorem}

\newtheorem{lemma}[theorem]{Lemma}

\theoremstyle{definition-boldhead}
\newtheorem{definition}{Definition}
\newtheorem{remark}{Remark}

\newdateformat{simple}{\THEDAY\ \monthname[\THEMONTH]\ \THEYEAR}
\simple

\floatstyle{ruled}
\newfloat{algo}{htbp}{algo}
\floatname{algo}{Algorithm}

\def \ifempty#1{\def\temp{#1} \ifx\temp\empty }
\renewcommand{\P}{\mathrm{P}}
\newcommand{\E}{\mathrm{E}}

\newcommand{\var}[1]{\textit{#1}}
\newcommand{\op}[1]{\textsl{#1}}

\newcommand{\false}{\textsc{false}\xspace}
\newcommand{\true}{\textsc{true}\xspace}
\newcommand{\etal}{\emph{et al.}}
\newcommand{\zo}{\ensuremath{\{0,1\}}\xspace}

\newcommand{\BN}{\ensuremath{\mathbb{N}}\xspace}

\newcommand{\BR}{\ensuremath{\mathbb{R}}\xspace}

\newcommand{\CL}{\ensuremath{\mathcal{L}}\xspace}

\newcommand\tx{\var{tx}\xspace}
\newcommand\txs{\var{txs}\xspace}
\newcommand\bl{\var{b}\xspace}
\newcommand\height{\op{height}\xspace}
\newcommand\gettxs{\op{txs}\xspace}
\newcommand\getcoms{\op{commitments}\xspace}
\newcommand\getminer{\op{miner}\xspace}
\newcommand\chain{\op{chain}\xspace}

\newcommand\bab{\op{bab}\xspace} % block-based atomic broadast
\newcommand\abbcast{\op{ab-broadcast}}
\newcommand\babbcast{\op{bab-broadcast}}
\newcommand\abdel{\op{ab-deliver}}
\newcommand\babdel{\op{bab-deliver}}
\newcommand\babmine{\op{bab-mined}\xspace}
\newcommand\data{\ensuremath{\var{data}\xspace}}

\newcommand\VT{\op{VT}\xspace}
\newcommand\VB{\op{VB}\xspace}
\newcommand\MB{\op{FB}\xspace}
\newcommand\ntx{\ensuremath{n_t }\xspace}

\newcommand\prot{\ensuremath{\Pi^3}\xspace}

\newcommand{\dotparagraph}[1]{\paragraph*{#1.}}

\usepackage{cleveref}

\usepackage[para]{footmisc}

\begin{document}

\title{\bf Eating sandwiches: Modular and lightweight elimination of transaction reordering attacks}

\author{Orestis Alpos\footnote{%
    Institute of Computer Science, University of Bern,
    Neubr\"{u}ckstrasse 10, 3012 CH-Bern, Switzerland.}\\
  University of Bern \\
  \url{orestis.alpos@unibe.ch}
  \and Ignacio Amores-Sesar\footnotemark[1]\\
  University of Bern\\
  \url{ignacio.amores@unibe.ch}
  \and Christian Cachin\footnotemark[1]\\
  University of Bern\\
  \url{christian.cachin@unibe.ch}
  \and Michelle Yeo\footnote{%
  IST Austria, Am Campus 1, 3400 Klosterneuburg, Austria.}\\
  IST Austria\\
  \url{michelle.yeo@ist.ac.at}
}

\date{\today}

\maketitle

\begin{abstract}\noindent
Traditional blockchains grant the miner of a block full control not only over which transactions but also their order.
This constitutes a major flaw discovered with the introduction of decentralized finance and allows miners to perform
MEV attacks. In this paper, we address the issue of sandwich attacks by providing a construction that takes as input a
blockchain protocol and outputs a new blockchain protocol with the same security but in which sandwich attacks are not
profitable. Furthermore, our protocol is fully decentralized with no trusted third parties or heavy cryptography
primitives and carries a linear increase in latency and minimum computation overhead.
\end{abstract}

\section{Introduction}
\label{sec:intro}
The field of blockchain protocols has proved to be extremely robust. Since its creation with Bitcoin~\cite{nakamoto2019bitcoin}, it had gone through several enhancements such as Ethereum~\cite{Ethereum} and has seen the appearance of \emph{decentralized finance} (DeFi). With this, some design flaws started to show up. Blockchains would ideally allow users to trade tokens with each other in a secure manner. However, existing designs do not consider users trading tokens of one platform for FIAT currency or tokens of a different platform, arguably one of the major flaws of today's blockchain platforms, \emph{maximal extractable value} (MEV)~\cite{DBLP:conf/sp/DaianGKLZBBJ20}. 
Current estimates show that the total volume of MEV since 2020 is around $675$M USD~\cite{mevflashbots}.
From a social welfare perspective, while MEV is profitable to miners, it presents a serious invisible tax on the users on the blockchain. Indeed the financial losses built up over time could potentially shy away users from the blockchain, and consequently impact the security of the chain. 

\emph{Sandwich attacks} are one of the most common types of MEV~\cite{KulkarniDC22} accounting for a loss of $174$M USD over the span of 33 months~\cite{DBLP:conf/sp/QinZG22} for users of Ethereum.
Sandwich attacks leverage the miner's ability to \emph{select} and \emph{position} transactions within a block. 
Consider the simple example of a sequence of transactions that swap one asset $X$ for another asset $Y$ in a decentralized exchange where exchange rates are computed automatically based on some function of the number of underlying assets in the pool (e.g., a constant product market maker~\cite{DBLP:conf/aft/AngerisC20}).
Now suppose there is a miner that also wants to swap some units of $X$ for $Y$.
The most favorable position for the miner would be to place their transaction at the start of the sequence, so as to benefit from a lower $X$-to- $Y$ exchange rate. This approach achieves a simple arbitrage strategy for any sequence of $X$-to-$Y$ swaps: the miner can insert an $X$-to-$Y$ exchange at the start of the sequence and use the computed exchange rate to sell, say, $k$ units of $X$ to get units of~$Y$. 
The miner then front-runs the sequence of $X$-to-$Y$ swaps, i.e., it inserts its own transaction at the start.
To finish off the attack, the miner back-runs the sequence with another transaction of its own that swaps some units of $Y$ to $X$, i.e., inserts this transaction at the end, and will often obtain more than $k$ units of~$X$.
In this way, the miner profits from its insider knowledge and its power to order transactions.
Refer to~\Cref{app:sandwich_detaildesc} for a detailed description of exchange rate computation and sandwich attacks.

Since any miner of a given block has full control over the transactions added to the block, as well as over the way transactions are ordered, it is straightforward for the miner to launch the above attack. 
Consequently, this gives miners a lot of power as they control precisely the selection and positioning of transactions with every block they mine. This problem has received broad attention in the practice of DeFi and in the scientific literature.

A classic technique to mitigate this attack is thus to remove the control over the positioning of the transactions in the block from the adversary, whether by using a trusted third party to bundle and order the transactions as in 
flashbots\footnote{\url{https://www.flashbots.net}}, Eden\footnote{\url{https://www.edennetwork.io}}, or OpenMEV\footnote{\url{https://openmev.xyz/}}.
Another method works by imposing a fair ordering of the transactions using a consensus algorithm that respects the order in which miners and validators first received the transactions~\cite{DBLP:conf/crypto/Kelkar0GJ20}. The classical solutions make the protocol dependent of external factors, besides, affecting efficiency.

In this work, we introduce the \emph{Partitioned and Permuted Protocol}, abbreviated $\Pi^3$, an efficient decentralized algorithm that does not rely on external resources to counter front-running. It renders sandwich attacks unprofitable and can easily be implemented on top of an existing blockchain protocol~$\Pi$.

Protocol~$\Pi^3$ determines the final order of transactions in a block $B_i$, created by a miner~$M_i$, through a \emph{uniformly randomly chosen permutation~$\Sigma_i$}.
To explain the method, let us focus on three transactions in $B_i$, a victim transaction $\var{tx}^*$
submitted by a client, and the front-running and back-running transactions, $\var{tx}_1$ and $\var{tx}_2$, respectively,
created by the miner. Since any relative ordering of these three transactions is equally probable, $\var{tx}_1$ will be ordered
before $\var{tx}_2$ with the same probability as $\var{tx}_2$ before $\var{tx}_1$,
hence the miner will profit or make a loss with the same probability.
Protocol~$\prot$ uses a fresh permutation for each block; it is chosen by a set of \emph{leaders},
which are recent miners in the blockchain.
We recognize and overcome the following challenges.

First, $\Sigma_i$ must not be known before creating $B_i$, otherwise $M_i$ would have the option to use $\Sigma^{-1}_i$,
the inverse of $\Sigma_i$, to initially order the transactions in $B_i$, so that the final order is the one that benefits $M_i$.
We overcome this by making $\Sigma_i$ known only \emph{after $M_i$ has been mined}.
On the other hand, if $\Sigma_i$ is chosen after creating $B_i$, a coalition of leaders would be able to try multiple different
permutations and choose the most profitable one~--- the number of permutations a party can try is only limited by their processing power.
For these reasons, we have the leaders \emph{commit} to their contributions to $\Sigma_i$ before $B_i$ becomes known, producing unbiased randomness. 
To incentivize leaders to open their commitments, our protocol $\Pi^3$ employs a delayed reward release mechanism that only releases the payment to leaders when they have generated and opened all commitments.

In some cases, however, performing a sandwich attack might still be more profitable than the block reward, and hence a leader might still choose to not reveal their commitment to bias the resulting permutation.
In general, a coalition of $k$ leaders can choose among $2^k$ permutations out of the $n_t !$ possible ones, where $n_t $ denotes the number of transactions in the block.
It turns out that the probability that $\var{tx}_1$, $\var{tx}^*$, and $\var{tx}_2$ appear in that order in one of the $2^k$ permutations can be significant for realistic values.
Protocol~\prot mitigates this by \emph{dividing each transaction} into $m$ chunks,
which lowers the probability of a profitable permutation in two ways.
First, the number of possible permutations is much larger, $(n_t m)!$ instead of $n_t !$. 
Second, a permutation is now profitable if the majority of chunks of $\var{tx}_1$ appear before the chunks of $\var{tx}^*$,
and vice versa for the chunks of $\var{tx}_2$.
As we discuss, the probability of a profitable permutation approaches zero rapidly as the number of chunks $m$ increases. We discuss how to implement the chunking mechanism while preserving transaction integrity and atomicity.

\dotparagraph{Organization}
In this paper, we introduce a construction that takes as input a blockchain protocol $\Pi$ and produces a new blockchain protocol $\Pi^3$ in which sandwich attacks are no longer profitable.
We begin by revisiting the concept of \emph{atomic broadcast}~\cite{DBLP:books/daglib/0025983} and setting the model for the analysis. Secondly, we introduce our construction justifying how miners are incentivized to follow the protocol before moving on to analyzing the construction in detail. Thirdly, we guarantee that the construction does not include any vulnerability to the protocol by showing that $\Pi^3$ implements a variant of atomic broadcast if $\Pi$ does. This part of the analysis is performed in the traditional \emph{Byzantine} model. Fourthly, we consider the \emph{rational} model to show that sandwich attacks are no longer profitable in $\Pi^3$. We consider the dual model of \emph{Byzantine} for the security analysis and \emph{rational} for the analysis of the sandwich attacks because we considered it to be a perfect fit to show that the security of $\Pi^3$ is not weakened even against an adversary that obtains nothing for breaking the protocol, as well as, we can assume that any party attempts to extract value from any sandwich attack. In other words, we consider both the security analysis and the analysis of the sandwich attack in the worst scenario possible for the protocol.
Lastly, we conclude the paper with an empirical analysis of the protocol under real-life data, as well as an analysis of the additional overhead introduced by our protocol.

\section{Related work}

The idea to randomize the transaction order within a block is folklore in the blockchain space.  It has been explored by Yanai~\cite{blinderswap} and also implemented in the wild~\cite{randomspam}. 
To the best of our knowledge, we are the first to implement the randomization using on-chain randomness and to provide a security analysis for this model. 
Additionally, Randomspam~\cite{randomspam} also acknowledges that some spamming attacks can occur with randomized transactions, where the attacker aims to insert several low-cost transactions to maximize the probability that some of these transactions are positioned exactly at a profitable transaction.
Our work reduces the success probability of these attacks by first chunking each transaction into smaller parts and then permuting all chunked transactions, rendering exact positioning attacks less profitable unless more transactions are added, incurring larger gas costs. 

\label{sec:related}
A recent line of work~\cite{DBLP:conf/crypto/Kelkar0GJ20, DBLP:conf/aft/Kursawe20, DBLP:conf/asiapkc/KelkarDK22, DBLP:conf/fc/CachinMSZ22, DBLP:journals/iacr/KelkarDLJK21}
formalizes the notion of \emph{fair ordering} of transactions.
These protocols ensure, at consensus level, that the final order is consistent with the local 
order in which transactions are observed by parties.
Similarly, the Hashgraph~\cite{DBLP:conf/coins/BairdL20} consensus algorithm aims to achieve fairness by having each
party locally build a graph with the received transactions. 
As observed by Kelkar~\etal~\cite{DBLP:conf/crypto/Kelkar0GJ20}, a transaction order
consistent with the order observed locally for any pair of transactions is not always possible, as Condorcet cycles may be formed.
As a result, fair-ordering protocols output a transaction order that is consistent with the view of only some \emph{fraction} of the 
parties, while some transactions may be output in a batch, i.e., with no order defined among them.
Moreover, although order-fairness removes the miner's control over the order of transactions,
it does not eliminate front-running and MEV-attacks: a \emph{rushing} adversary that becomes aware
of some $\var{tx}$ early enough can broadcast its own $\var{tx}'$ and make sure that sufficiently many nodes receive $\var{tx}'$ before $\var{tx}$.

Another common defense against front-running attacks is the \emph{commit and reveal} technique.
The idea is to have a user first commit to a transaction, e.g., by announcing its hash or its encryption,
and, once the order is fixed, reveal the actual transaction.
However, an adversary can choose not to reveal the transaction, should the final order be non-optimal.
Doweck and Eyal~\cite{DBLP:journals/corr/abs-2005-04883} employ time-lock puzzle commitments~\cite{time-lock}, so that a transaction can be brute-force revealed, and protocols such as Unicorn~\cite{DBLP:journals/ijact/LenstraW17} and Bicorn~\cite{DBLP:journals/iacr/ChoiATB23} employ verifiable delay functions~\cite{DBLP:conf/crypto/BonehBBF18} to mitigate front-running.
Whereas they indeed manage to mitigate front running, the main disadvantages of these solutions are threefold: firstly, transactions may be executed much later than submitted, with no concrete upper bound on the revelation time. Secondly, a delay for the time-lock puzzle has to be chosen which matches the network delay and adversary's computational power. Finally, it is unclear who should spend the computational power to solve the time-lock puzzles, especially in proof of work blockchains where this shifts computational power away from mining.

A different line of work~\cite{DBLP:conf/dsn/DuanRZ17, DBLP:conf/ccs/MillerXCSS16, DBLP:conf/crypto/CachinKPS01, DBLP:journals/toplas/ReiterB94,F3B} hides the transactions until they are ordered with the help of a \emph{committee}.
For instance, transactions may be encrypted with the public key of the committee, so that its members can collaboratively decrypt it.
However, this method uses threshold encryption~\cite{DBLP:conf/crypto/DesmedtF89} and requires a coordinated setup.
Also multi-party computation (MPC) has been used~\cite{DBLP:conf/acns/BaumDF21, DBLP:conf/ccs/AbrahamPY20,DBLP:conf/ccs/LuYKGKM19}
to prevent front-running. MPC protocols used in this setting must be tailor-made so that misbehaving is identified and punished~
\cite{DBLP:conf/fc/BaumDD20, DBLP:conf/eurocrypt/KiayiasZZ16}.
A disadvantage of the aforementioned techniques is that the validity of a transaction can only be checked after it is revealed. These techniques also rely on strong cryptographic assumptions and coordination within the committee.
The protocol presented in this work disincentivizes sandwich attacks without requiring hidden transactions or
employing computationally heavy cryptography.

Another widely deployed solution against front-running involves a dedicated \emph{trusted third party}.
Flashbots\footnote{\url{www.flashbots.net}}, Eden\footnote{\url{www.edennetwork.io}},
and OpenMEV\footnote{\url{https://openmev.xyz/}} 
allow Ethereum users to submit transactions to their services, then order received
transactions, and forward them to Ethereum miners.
Chainlink's Fair Sequencing Service~\cite{chainlink2}, in a similar fashion, aims to collect encrypted transactions from users,
totally orders them, and then decrypts them. The third-party service may again be run in a distributed way.
The drawback with these solutions is that attacks are not eliminated, but trust is delegated to a different set of parties.

An orthogonal but complementary line of research is taken by Heimbach and Wattenhofer~\cite{DBLP:conf/asiaccs/HeimbachW22}.
Instead of eliminating sandwich attacks, they aim to improve the resilience of ordinary transactions against sandwich attacks by strategically setting their slippage tolerance to reduce the risk of both transaction failure as well as sandwich attacks.

Last but not least, Baum~\etal~\cite{DBLP:journals/iacr/BaumCDFG21} and Heimbach and Wattenhofer~\cite{DBLP:conf/asiaccs/HeimbachW22} survey the area of front-running attacks.

\section{Model}\label{sec:basics}
\dotparagraph{Notation}
For a set $X$, we denote the set of probability distributions on $X$ by $\mu(X)$.
For a probability distribution $\nu\in\mu(X)$, we denote sampling $x$ from $X$ according to $\nu$ by $x \gets \nu$.

\subsection{Block-based atomic broadcast}

Parties broadcast transactions and deliver blocks using the events 
$\babbcast(\tx)$ and $\babdel(\bl)$, respectively,
where block \bl contains a sequence of transactions $[\tx_1, \ldots, \tx_{\ntx}]$.
The protocol outputs an additional event $\babmine(\bl, P)$, 
which signals that block~\bl has been \emph{mined} by party $P$, where $P$ is defined as the \emph{miner} of \bl.
Notice that $\babmine(\bl, P)$ signals only the creation of a block and not its delivery.
In addition to predicate $\VT()$, we also equip our protocol
with a predicate $\VB()$ to determine the validity of a block.
Moreover, we define a function $\MB()$, which describes how to fill a block:
it gets as input a sequence of transactions and any other data required by the protocol and outputs a block.
These predicates and function are determined by the higher-level application or protocol.

\begin{definition}\label{def:bab}
  A protocol implements \emph{block-based atomic broadcast} with validity predicates $\VT()$ and $\VB()$ and block-creation function $\MB()$ if it satisfies the following properties, except with negligible probability:
 \begin{description}
  \item[Validity:] If a correct party invokes a $\babbcast(\tx)$, then every correct party eventually outputs $\babdel(\bl)$, for some block \bl that contains $\var{tx}$.
  \item[No duplication:] No correct party outputs $\babdel(\bl)$ for a block $\bl$ more than once.
  \item[Integrity:] If a correct party outputs $\babdel(\bl)$, then it has previously output the event $\babmine(\bl, \cdot)$ exactly once.
  \item[Agreement:] If some correct party outputs $\babdel(\bl)$, then eventually every correct party outputs $\babdel(\bl)$.
  \item[Total order:] Let $\bl$ and $\bl'$ be blocks, and $P_i$ and $P_j$ correct parties that output $\babdel(\bl)$ and $\babdel(\bl')$. If $P_i$ delivers \bl before $\bl'$, then $P_j$ also delivers $\bl$ before $\bl'$.
  \item[External validity:] If a correct party outputs $\babdel(\bl)$,
  such that $\bl=[\tx_1, \ldots, \tx_{\ntx}]$,
  then $\VB(\bl)=\true$ and $\VT(\tx_i) = \true$, for $i \in 1, \ldots, \ntx$.
  Moreover, if $\MB(\tx_1, \ldots, \tx_{\ntx})$ returns \bl, then $\VB(\bl)=\true$.
  \item[Fairness:] There exists $C \in \BN$ and $\mu \in \BR_{>0}$, such that for all $N\geq C$ consecutive delivered blocks,
  the fraction of the blocks whose miner is correct is at least $\mu$.
 \end{description}
\end{definition}

Observe that the properties assure that $\babmine(\bl, P)$ is triggered exactly once for each block \bl,
hence each block has a unique miner.
For ease of notation, we define on a block \bl the fields $\bl.\gettxs$,
which contains its transactions, and $\bl.\getminer$, which contains its miner.
Since blocks are delivered in total order, we can assign them a \emph{height}, a sequence number in their order of delivery,
accessible by $\bl.\height$. Finally, for simplicity we assume that a delivered block allows access to all blocks with smaller
height, through an array $\bl.\chain$. That is, if $\bl.\height = i$ then $\bl.\chain[i']$ returns $\bl'$, such that
$\bl'.\height = i'$, for all $i' \leq i$.

\subsection{Blockchain and network}
\label{sec:model}
Blockchain protocols derive their security from different techniques such as \emph{proof of work} (\emph{PoW})~\cite{nakamoto2019bitcoin}, \emph{proof of stake} (\emph{PoS})~\cite{DBLP:conf/eurocrypt/DavidGKR18}, \emph{proof of space-time} (\emph{PoST})~\cite{chia}, or \emph{proof of elapsed time} (\emph{PoET})~\cite{DBLP:conf/indocrypt/BowmanDMM21}. 
In the remainder of the work, we consider a generic protocol $\Pi$ that has a probabilistic termination condition, capturing all the model above. Furthermore, we model $\Pi$ as block-based atomic broadcast.

\dotparagraph{Parties}
Similar to previous works, our protocol does not make explicit use 
of the number of parties or their identities, and does not require the parties
themselves to know this number. We assume an static network of $n_\var{p}$ parties.
We consider the Byzantine model, where $f$ parties may behave arbitrarily, as well as, the rational model where all parties behave maximizing their utilities.

\dotparagraph{Transactions \& Blocks}
A transaction~$\var{tx}$ contains a set of \emph{inputs}, a set of \emph{outputs},
and a number of digital signatures. 
Transactions are batched into \emph{blocks}.
A block contains a number of transactions, $n_t $, for simplicity we assume  $n_t $ to be constant.
A block \bl may contain parameters specific to protocol~$\Pi$  such as references to previous blocks, but we abstract the logic of accessing them in a field $\bl.\chain$, as explained in the context of Definition~\ref{def:bab}. We allow conditional execution of transactions across blocks, i.e., a transaction can be executed conditioned on the existence of another transaction in a previous block.

\dotparagraph{Network}
A \emph{diffusion functionality} implements communication among the parties, which is structured into \emph{synchronous} rounds.
The functionality keeps a $\op{RECEIVE}_i$ string for each party $P_i$ and makes it available to $P_i$ at the start of every round.
String $\op{RECEIVE}_i$ is used to store all messages $P_i$ receives.
When a party $P_i$ instructs the diffusion functionality to \emph{broadcast} a message, we say that
$P_i$ has \emph{finished its round} and the functionality tags $P_i$ as finished for this round.
The adversary, detailed in Section~\ref{sec:analysis}, is allowed to read the string of any party at any moment during the execution and to see any messages broadcast by any party immediately. Furthermore, the adversary can write messages directly and selectively into $\op{RECEIVE}_i$ for any $P_i$, so that only $P_i$ receives the message at the beginning of the next round. This models a \emph{rushing} adversary.

When all non-corrupted parties have finished their round, the diffusion functionality takes all messages that were broadcast by non-corrupted parties in the round and adds them to $\op{RECEIVE}_i$ for all parties, this is the reason of the name \emph{synchronous} rounds. Every non-corrupted party communicates changes to its local view at the end of each round. If a non-corrupted party creates a block in round~$r$, the new block is received by all parties by round $r+1$.
Furthermore, even if the adversary causes a block to be received selectively by only some non-corrupted parties in round $r$, the block is received by all non-corrupted parties by round $r+2$. The update of the local view also includes the delivery of transactions contained in the blocks that satisfy the conditions to be accepted.

\section{Protocol}\label{sec:protocol}

Our proposed protocol $\Pi^3$ (``Partitioned and Permuted Protocol'') contains two modifications to a given underlying blockchain protocol $\Pi$ in order to prevent sandwich MEV attacks.

Our first modification involves randomly permuting the transactions in any given block. 
Note that a naive way of doing so is to use an external oracle (e.g., DRAND~\cite{Drand} or NIST beacon~\cite{nist}) to generate the randomness which will be applied to a given block.
However, using an external source of randomness relies on strong trust assumptions on the owners of the source, leaves our protocol vulnerable to a single point of failure and it introduces incentive issues between miners of the chain and owners of the external source.
To avoid this, our protocol uses miners of the immediately preceding blocks to generate the randomness.
These miners, which we refer to as \emph{leaders} for the given block, are in charge of generating random \emph{partial seeds}.
These partial seeds are then combined to form a \emph{seed} which will be the input into a PRG to produce a random permutation that is applied to the transactions in the block.
To ensure that the permutation is random, we need first to achieve that leaders participate in the generation of random partial seeds and secondly to ensure the partial seeds generated by the leaders are random. 
That is, the leaders should not commit to the same partial seed each time or collude with other leaders to generate biased partial seeds. 
To incentivize each leader to participate in the generation of the seed, $\Pi^3$ stipulates that they commit to their partial seed and present a valid opening during the commitment opening period, otherwise their reward will be burned. 
In typical blockchain protocols, the miner of a block receives the block reward immediately.
In $\Pi^3$, the miner does not receive the reward until a certain number of additional blocks has been mined. 
We refer to this as a \emph{waiting phase} and stress that the precise length of the waiting phase is a parameter in our protocol that can be tweaked.

Our second modification is to divide the transfers of each transaction into smaller chunks before permuting the chunked transactions of a block. 
This modification increases the cardinality of the permutation group in order to reduce the effectiveness of any attack aiming to selectively open partial seeds in order to bias the final permutation.

We stress that our proposed modifications incur minimal computational overhead, since the only possible overhead corresponds to transaction delivery and this aspect is computationally cheap. Thus, the only noticeable impact of our protocol is latency. 
In~\Cref{sec:efficiency_analysis} we provide an in-depth analysis of the efficiency impact of our proposed modifications.

\subsection{Permuting transactions}
\label{sec:protocol-details}

\begin{figure}
  \centering
  \includegraphics[width=\textwidth]{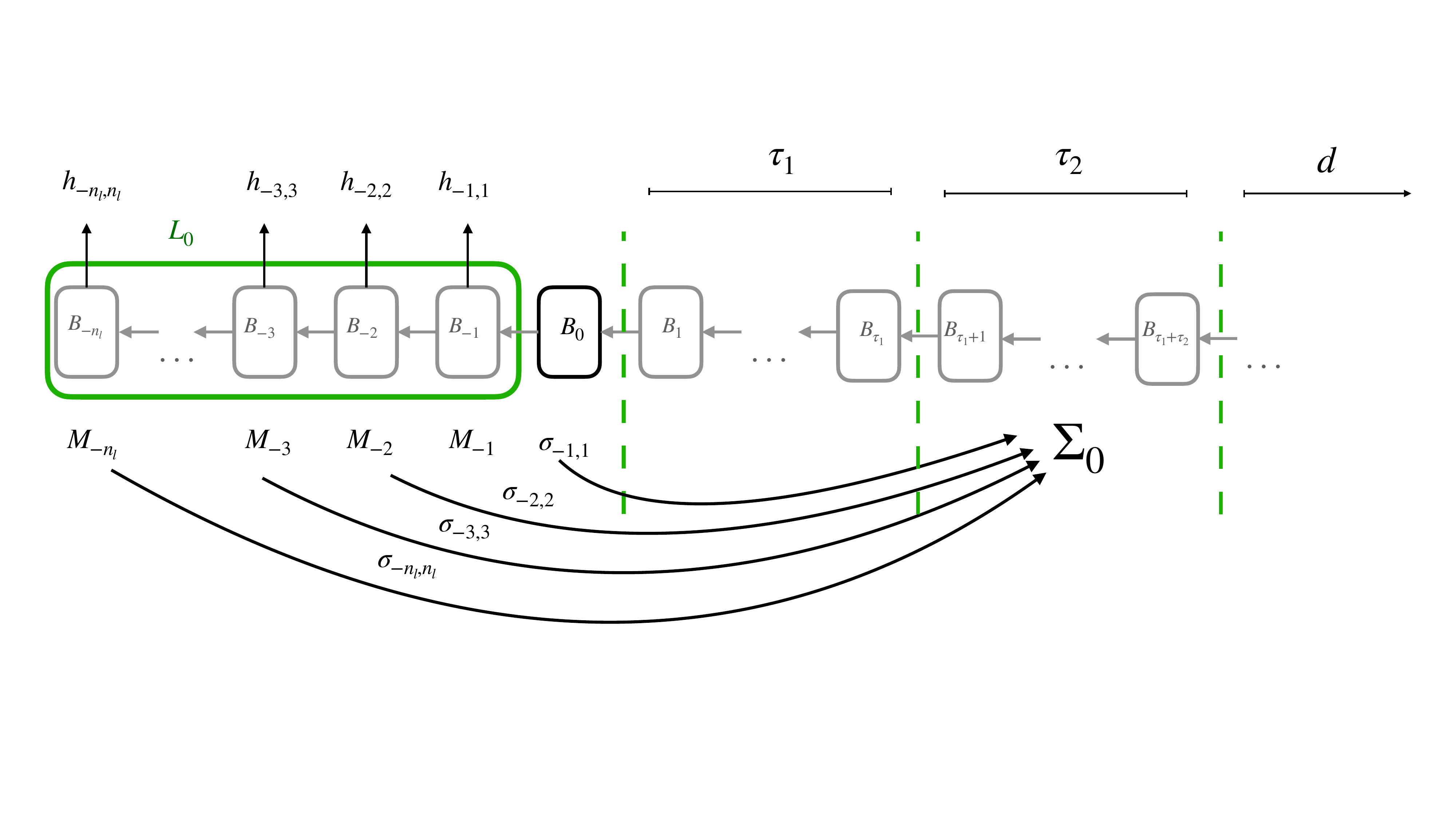}
  \vspace{-65 pt}
  \caption{We illustrate the routine for the creation of the random permutation $\sigma_0$ created by the leader set $L_0$ and to be applied on block $B_0$ in Bitcoin. The leaders $L_0$ for block $B_0$ is formed by the miners of the $n_\ell$ blocks before $B$, marked with the green box. When party $M_i$ mined block $B_i$, the party generated some random seed $\sigma_{-i,i}$ and included its commitment $h_{-i,i}$ as part of the newly mined block. After block $B_0$ is mined, the leaders wait for $\tau_1$ blocks before opening the commitments. The commitments must be included in the following $\tau_2$ blocks. Finally the parties wait until every block containing openings are confirmed before delivering block $B_0$}   
 \label{fig:example}
\end{figure}

Protocol $\Pi^3$ consists of the following four components (see~\Cref{fig:example}): block mining, generation of the random permutation, reward (re)-distribution, and chunking the transactions. 

\label{sec:perm_from_seed}

\dotparagraph{Appending the partial seeds}
\label{sec:hashgood}
Let $n_\ell$ be the size of the leader set for each block. 
The miner $M_i$ of block $B_i$ is part of the leader set of blocks $B_{i+j}$, for $j \in [n_\ell]$.
$M_i$ must therefore contribute a partial seed $\sigma_{i,j}$ for each of these $n_\ell$ blocks following $B_i$.
Hence, $M_i$ needs to create $n_\ell$ random seeds $\sigma_{i, 1}, \ldots, \sigma_{i, n_\ell}$ and commitments to them, $C(\sigma_{i, 1}), \ldots, C(\sigma_{i, n_\ell})$.
The commitments $C(\sigma_{i, j})$, for $j \in [n_\ell]$, are appended to block $B_i$, while the seeds $\sigma_{i, j}$ are stored locally by $M_i$.
A block that does not contain $n_\ell$ commitments is considered invalid.

Looking ahead, we want that any party knowing the committed value can demonstrate it to any other party. Thus, the more standard commitments schemes such as Pedersen commitment~\cite{DBLP:reference/crypt/Pedersen05} are ill-suited. Instead, $\Pi^3$ uses a deterministic commitment scheme for committing to permutations, in particular, a collision-resistant cryptographic hash function.
When the entropy of the committed values is high enough, then a hash function constitutes a secure commitment scheme. Since the parties commit to a random partial seed, hash functions suffice and yield a cheap commitment scheme.

\dotparagraph{Opening the commitments}
Let $\tau_1, \tau_2 \in \mathbb{N}_{> 0}$. 
Between $\tau_1$ and $\tau_1+\tau_2$ blocks after the creation of some block $B_i$, the commitments of the partial permutation to be applied on block $B_i$ must be opened. 
The miners of these blocks also need to append the openings to their blocks, unless a previous block in the chain already contains them (see below for more details). 
The parameter $\tau_1$ controls the probability of rewriting block $B_i$ after the commitments have been opened.
Whereas, parameter $\tau_2$ guarantees that there is enough time for all the honest commitments to be opened and added to some block. Any opening appended a block $B_j$ for $j > i+\tau_1 + \tau_2$ is ignored. 
We note that specific values of $\tau_1$ and $\tau_2$ might cause our protocol to suffer an increase in latency. 
We leave these parameters to be specified by the users of our protocol. For the interested reader, we discuss latency-security trade-offs in Section~\ref{sec:eth-analysis}. The $\tau_1$ blocks created until opening the commitments takes place is known as \emph{silent phase}, whereas the following $\tau_2$ blocks is known as \emph{loud phase}.

A possible way to record the opening of commitments is for the miners that own the commitments to deploy a smart contract that provides a method $\op{open}(i,j,\sigma_{i,j})$,
where $\sigma_{i,j}$ is a (claimed) opening of the $j$-th commitment $h_{i,j}$
published in the $i$-th block $B_i$.
We remark, that the smart contract serves only as proof that an opening
to a commitment has been provided, and does not add any functionality to the protocol, so other proof mechanisms can also be considered.
The protocol monitors the blockchain for calls to this method. 
The arguments to each call, as well as the calling party and the block it appears on, are used to determine the final permutations of the blocks and the distribution of the rewards, which we will detail below. We stress that \emph{not} opening a commitment does not impact the progress of protocol, as unopened commitments are ignored.

\dotparagraph{Deriving the permutation from partial seeds}
Let the seed $\sigma_i$ for block $B_i$ be defined as $\sigma_{i-1,1} \oplus \sigma_{i-2,2} \oplus \ldots \oplus \sigma_{i-n_\ell, n_\ell}$.
Given the seed $\sigma_i$, let $r_i := \op{G}(\sigma_i)$, where $G: \{ 0,1 \}^\lambda \rightarrow \{ 0,1 \}^\ell$
is a pseudorandom generator. If at least one of the partial seeds $\sigma_{i,j}$, for $j \in [n_\ell]$,
is chosen at random, then $\sigma_i$ is random as well, and $r_i$ is indistinguishable from a random number~\cite{phone} without the knowledge of $\sigma_{i,j}$.
There are standard algorithms to produce a random permutation from a polynomial number of bits~\cite{DBLP:books/daglib/0023376}.

\dotparagraph{Incentivizing the behavior}\label{sec:reward}
A crucial factor in the security of $\Pi^3$ against sandwich MEV attacks is that the permutation used to order transactions within a block should be truly random. 
Thus, the miners should generate all partial seeds uniformly at random. 
To incentivize them to do so, we exploit the fact that all leaders remain in the waiting phase for a period of time, which means that they have not yet received the block rewards and fees for mining their block on the blockchain. Note that the waiting phase is $n_\ell+\tau_1+\tau_2+d$ blocks long.
This implies that their rewards can be claimed by other miners or burned if a party diverges from the proper execution, according to the rules described below. Consider a partial permutation $\sigma_{i,j}$ committed by miner $M_i$ of block $B_i$. Recall that $\sigma_{i,j}$ will be applied on block $B_{i+j}$ and that miners can be uniquely identified due to the $\babmine()$ event.

\begin{enumerate}
    \item Before $\tau_1$ blocks have been appended after block $B_{i+j}$ any other leader of the leader set $\CL_{i+j}$ who can append a pre-image of $h_{i,j}$ to the chain can receive the reward and fees corresponding to $M_i$. This mechanism prevents party $M_i$ from disclosing its commitment before every other leader committed its randomness, thus preventing colluding. A miner whose commitment has been discovered by another leader is excluded from all the leader sets. 

    \item If the opening of $\sigma_{i,j}$ is not appended to any block, miner $M_i$ loses its reward and fees. This mechanism prevents miners from not opening their commitments. Note that miners are incentivized to include all the valid openings, as discussed below.

    \item  If any of the previous conditions do not apply, party $M_i$ receives an $\alpha$ fraction of the block reward and feed for $\alpha \in (0,1)$, which would be paid out the moment $M_i$ leaves the waiting phase.
    Each miner that appends the opening of $M_i$'s commitments gets $\frac{(1-\alpha) \cdot w}{n_\ell}$ for each commitment appended.
\end{enumerate}
In the remainder of this work we refer to the block reward and fees as simply \emph{block reward}.

\subsection{Chunking transactions}\label{sec:chunk}

\begin{figure}
  \centering
  \includegraphics[width=0.9\textwidth]{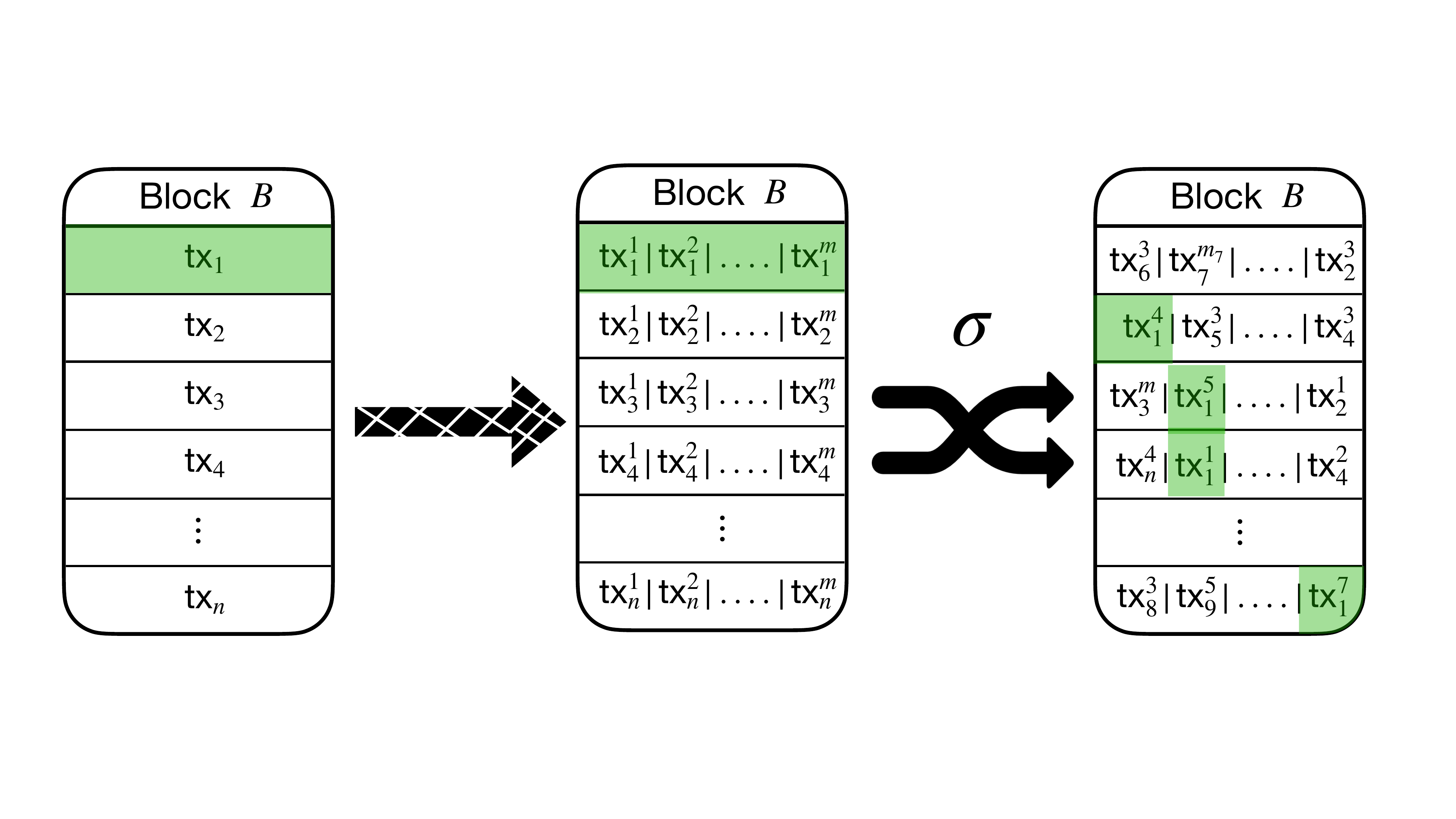}
  \vspace{-40 pt}
  \caption{We illustrate the process followed by party $M$ to deliver a block $B$. We denote by $\var{tx}_1,..,\var{tx}_n$ the $n$ transaction that constitute $B$. In the first step, party $M$ breaks each transaction $\var{tx}_i$ into $m$ transactions $\var{tx}_i^1,..,\var{tx}_i^{m}$ involving a smaller amount. These smaller transactions are later permuted according to the random permutation $\Sigma$. Lastly, party $M$ delivers these small transactions in this new order.}
\end{figure}

In all commit-and-open schemes, there exists the vulnerability that malicious parties decide to not open their commitments so as to bias the outcome. 
In our protocol, any coalition of $k$ leaders can choose between $2^k$ ways to bias the final permutation.
If one miner manages to create multiple blocks out of $B_{i-1}, \ldots, B_{i-n_\ell}$, this does not even require collusion with others.

The adaptive attacks mounted through withholding can be countered with a
simultaneous broadcast abstraction~\cite{DBLP:conf/focs/ChorGMA85}, but
realizing this is almost impossible in
practice~\cite{DBLP:conf/stoc/KolN08}, especially in the blockchain domain.
Alternatively, time-lock puzzles may negate the effect of the delay.  But
this technique costs computational effort, which may have a negative impact
on the environment and possibly also on the protocol.  In the particular
case of generating a permutation, there is an alternative.

\dotparagraph{Chunking transfers in transactions}

Let us assume that a block contains $n_{\var{tx}}$ transactions, this means there exist $n_{\var{tx}}!$ possible permutations of them. A coalition of $k$ leaders can choose between $2^k$ possible permutations among the $n_{\var{tx}}!$ total permutations. Furthermore, in the simplest case the coalition only aims to order the three transactions that constitute the sandwich attack, thus the fraction of advantageous permutations is $\frac{1}{6}$, the fraction of disadvantageous permutations is $\frac{1}{6}$ and the remaining ones are neutral. If $k$ is big enough, the coalition could still extract enough value to compensate for the lost block rewards of those parties that do not open their commitment.

Therefore we want to increase the size of the permuted space.  We assume here that every transaction consists of arbitrary code and a few specialized instructions that are \emph{transfers} of coins or tokens.  These may be the \emph{native payment operation} of the blockchain or  operations that involve a well-known \emph{standard format for tokens}, which are emulated by smart contracts (such as the ERC-20 standard in Ethereum). We now divide every transfer generated by some transaction into $m$ chunks.

For instance, suppose transaction $\var{tx}_i$ consists of Alice paying Bob 1 ETH. In our protocol, each party would locally split $\var{tx}_i$ into $m$ chunks $\var{tx}_i^1, \dots \var{tx}_i^m$, consisting of Alice transferring $1 / m$ ETH to Bob each.
After all transactions are chunked, the permutation will be applied to the larger set of transactions. There exist $(n_{\var{tx}} m)!$ permutations and the coalition would need to order the $3m$ chunks that constitute the involved transactions. Furthermore, for a given permutation with some chunks ordered beneficially, there will exist chunks ordered in a disadvantageous way, with overwhelming probability. The coalition needs to optimize the good ordering of some chunks while keeping the bad ordering under control. Obtaining a favorable ordering becomes extremely unlikely as number of chunks $m$ grows (Section~\ref{sec:game_analysis}).

\dotparagraph{Execution of transactions and chunks}
Transactions contain arbitrary code whose execution produces an ordered
list of transfers, as introduced before.  The process of chunking proceeds
in two stages.

In the first stage, the party executes the code of all transactions
contained in the block serially, in the order determined by the miner; this
produces a list of transfers and the corresponding amounts for each
transaction.  Some transactions may turn out to be invalid, they are
removed from the further processing of the block.

In the second stage, for each valid transaction \var{tx} a list of $m$ transfer
chunks $\var{tx}^1, \dots \var{tx}^m$ is produced such that $\var{tx}^1$
contains the \emph{code} executed by \var{tx}, and its transfers
have their amount set to $1/m$ of the number computed by the code.
The transaction
chunks $\var{tx}^2, \dots \var{tx}^m$ contain \emph{only} the transfers,
with their numbers set to $1/m$ of original amounts, but these transactions do not execute
further code. If the code executed by $\var{tx}^1$ produces different transfers than in the first stage,
the execution of $\var{tx}^1$ is aborted, also the execution of 
$\var{tx}^2, \dots \var{tx}^m$.
We consider two transfers to be the same if they transfer the
same amount of coins or tokens from the same source address to the same destination address. Note 
the blockchain state produced by a transaction in the first stage can differ
from the state produced by the same transaction in the actual execution
according to the second stage execute; this is the case, for instance,
when it interacts with a smart contract.

The block being permuted now contains up to $m$ times as many
chunks as the original block contained transactions, each of them
transferring $1/m$ the value.

Notice that the permutation is not uniformly random across all choices,
but needs to respect that $\var{tx}^1$ appears first within the set of
chunks resulting from \var{tx}. However, this restriction in the permutation
does not constitute any loss of generality since every chunk performs an
identical transfer.
An adversarial miner can  utilize fine-grained conditions such as slippage to additionally control the conditional execution of transactions~-- and in our case transaction chunks~-- in a given block. The execution of transactions explained above guarantees atomicity: all chunks are executed or no chunk is executed. In~\Cref{sec:sandwich_randomperm} we present an in-depth analysis of how slippage could lead to higher expected revenue, which may also be of independent interest.

\subsection{Details}

\begin{algo*}
  \vbox{
  \small
  \begin{numbertabbing}
    xxxx\=xxxx\=xxxx\=xxxx\=xxxx\=xxxx\=MMMMMMMMMMMMMMMMMMM\=\kill
    \textbf{Implements:} Protocol~$\Pi^3$ \\
    \textbf{Uses:} block-based atomic broadcast~$\Pi$ \\
    \\
    \textbf{State:}\\
    \> $\sigma[i,j] \gets \bot$, for all $i \geq 1, j \in [n_\ell]$ \label{}\\
    \> $c[i,j] \gets \bot$, for all $i \geq 1, j \in [n_\ell]$ \label{}\\
    \\
    \textbf{upon event} $\langle \Pi^3\op{-broadcast}, \tx \rangle$ \textbf{do}\label{line:bcast-begin}\\
    \>\textbf{invoke} $\langle \Pi\op{-broadcast}, \tx \rangle$ \label{line:bcast-end}\\
    \\
    \textbf{upon event} $\langle \Pi{-mined}, \bl, Q \rangle$ \textbf{do}\label{line:babmine-begin}\\
    \> $i_{\text{open}} \gets \bl.\height - \tau_1 - 1$ \label{line:block-to-open}\\
    \>\textbf{for} $i' \in [i_\text{open} - n_\ell - 1, i_\text{open} - 1]$ \textbf{do}\label {line:loop-through-mined-begin}\\
    \>\> \textbf{if} $\bl.\chain[i'].\op{miner} = P$ \textbf{then} \label{}\\
    \>\>\>$\op{Open}(\bl.\chain[i'].\getcoms[i_\text{open} - i'])$ \label{line:loop-through-mined-end}\\

    \\
    \textbf{upon event} $\langle \Pi\op{-deliver}, \bl \rangle$ \textbf{do}\label{line:babdel-begin}\\
    \> $i_{\text{del}} \gets \bl.\height - \tau_1 - \tau_2$ \label{line:delivered-block}\\
    \> \textbf{for} $j \in [n_\ell]$ \textbf{do} \label{line:com-extraction-begin} \`// Read commitments for block~\bl \\
    \>\> $c[i_\text{del},j] \gets \bl.\chain[i_\text{del} - j].\getcoms[j]$ \label{line:com-extraction-end}\\
    \>\textbf{for} $i' \in [i_\text{del} + \tau_1 + 1, i_\text{del} + \tau_1 + \tau_2]$ \textbf{do}\label {line:opening-extraction-begin} \`// Read the openings for block~\bl \\
    \>\>\textbf{for} $\tx \in  \bl.\chain[i'].\gettxs$  \textbf{such that} $\tx = \op{open}(k, l, \sigma)$ \textbf{do}\label{}\\
    \>\>\> \textbf{if} $k + l = i_\text{del}$  \textbf{and} $H(\sigma) = c[i_\text{del}, l]$ \textbf{then} \label{line:opening-checks}\\
    \>\>\>\> $\sigma[i_\text{del}, l] \gets \sigma$ \label{line:opening-extraction-end}\\
    \> $\var{seed} \gets 0^\lambda$ \label{line:derive-permutation-begin}\\
    \> \textbf{for} $j \in [n_\ell]$ \textbf{do} \label{}\`// Compute final permutation for block~\bl\\
    \>\> \textbf{if} $\sigma[i_\text{del}, j] \neq \bot$ \textbf{then} \label{}\\
    \>\>\> $\var{seed} \gets \var{seed} \oplus \sigma[i_\text{del}, j]$ \label{}\\
    \> $\Sigma \gets \op{PermFromRandBits}(G(\var{seed}))$ \label{line:derive-permutation-end}\\

    \> $\var{chunked\_txs} \gets [\,]$ \label{line:chunk-begin}\\
    \>\textbf{for} $\tx \in \bl.\gettxs$ \textbf{do} \label{}
     \`// Chunk and permute transactions in block~\bl\\
    \>\> $\var{chunked\_txs} \gets \var{chunked\_txs} \| \op{Chunk}(\tx, m)$ \label{line:chunk-end}\\
    \> $\var{chunks} \gets \op{Permute}(\Sigma,\var{chunked\_txs})$\label{line:permute}\label{}\\
    \> $\var{chunks}\gets \op{SwapChunks}(\var{chunks})$\label{line:swap}\`// For each $\var{tx}$, swap the first chunk in $\var{chunks}$ with $\var{tx}^1$ \\
    \>$\bl.\gettxs \gets \var{chunks} $\label{}\\
    \>\textbf{invoke} $\langle \Pi^3\op{-deliver}, b \rangle$ \label{line:deliver-end}\\
    \\
    \textbf{function} $\MB(\txs)$ \textbf{:} \label{}\\
    \> $\data \gets [\,]$ \label{}\\
    \> \textbf{for} $\tx \in \txs$ \textbf{do} \label{}\\
    \>\> $\data \gets \data \| tx $ \label{}\\
    \> \textbf{for} $j \in [n_\ell]$ \textbf{do} \label{}\\
    \>\> $\sigma \stackrel{\$}{\gets} \zo^\lambda$\label{}\\
    \>\> $c \gets H(\sigma)$ \label{}\\
    \>\> $\data \gets \data \| c$ \label{}\\
    \> \textbf{return} $ \data$ \label{}\\
    \\
    \textbf{function} $\VB{}(\bl)$ \textbf{:} \label{}\\
    \>\textbf{if} $\bigl(\exists \tx \in \bl.\gettxs: \neg\VT(\tx) \bigr)
      \lor \bigl(\exists j \in [n_\ell]: \bl.\getcoms[j] = \bot\bigr)$ 
      \textbf{then} \label{}\\
    \>\> \textbf{return} \false \label{}\\
    \> \textbf{return} \true \label{}
    
  \end{numbertabbing}
  }
  \caption{Protocol~\prot. Code for party~$P$.}
  \label{algo:functionality}
\end{algo*}

In Algorithm~\ref{algo:functionality} we show the pseudocode for protocol~\prot,
which implements a block-based atomic broadcast (\bab) primitive.
The pseudocode assumes an underlying protocol $\Pi$, which is also modeled as a block-based atomic broadcast (\bab) primitive,
as defined in Section~\ref{sec:basics}.
The user or high-level application interacts with \prot by invoking $\Pi^3\op{-broadcast}(\tx)$ events. These are handled
by invoking the corresponding $\Pi\op{-broadcast}(\tx)$ event on the underlying protocol $\Pi$ (L\ref{line:bcast-begin}-\ref{line:bcast-end}).

Protocol~$\Pi$ outputs an event $\babmine(\bl, Q)$ whenever some party $Q$ mines a new block \bl (L\ref{line:babmine-begin}).
For \prot, the mining of a new block at height $i$ starts the opening phase for the block at height
$i_\text{open} = i - \tau_1 - 1$ (L\ref{line:block-to-open}).
Hence, party~$P$ loops through the $n_\ell$ blocks before $i_\text{open}$ and checks whether it is the miner
of each of them (L\ref{line:loop-through-mined-begin}-\ref{line:loop-through-mined-end}).
If this is the case, $P$ must provide a valid opening to the commitment related to block at height $i_\text{open}$.
The opening is achieved by a specific type of transaction, for example through a call to a smart contract.
In the pseudocode we abstract this into a function $\op{Open}()$.

Protocol~$\Pi$ outputs an event $\Pi\op{-deliver}(\bl)$ whenever a block \bl is delivered (L\ref{line:babdel-begin}).
According to the analysis of our protocol, this will allow \prot to deliver the block 
$\tau_1 + \tau_2$ positions higher than \bl, i.e., the block $\bl_\text{del}$ at height $i_{\text{del}} = \bl.\height - \tau_1 - \tau_2$.
To this goal, \prot first reads the commitments related to $\bl_\text{del}$ (L\ref{line:com-extraction-begin}-\ref{line:com-extraction-end}).
By construction of \prot, a commitment $c_{i,j}$, written on block $\bl_i$, is used to order the
transactions in block $\bl_{i+j}$.
Hence, the commitments related to $\bl_\text{del}$ have been written on the $n_\ell$ blocks before $\bl_\text{del}$.
Protocol \prot then reads the openings to these commitments (L\ref{line:opening-extraction-begin}-\ref{line:opening-extraction-end}). Again by construction of \prot,
the openings of
the commitments related to $\bl_\text{del}$ have been written on the blocks with height $i_\text{del}+\tau_1 + 1$ to
$i_\text{del}+\tau_1 + \tau_2$. For each of these blocks, \prot loops through its transactions that contain an opening.
L\ref{line:opening-checks} then checks whether the opening is for a commitment related to block $\bl_\text{del}$
and whether the opening is valid. 
Protocol \prot then calculates the final permutation $\Sigma$ to be applied to block \bl (L\ref{line:derive-permutation-begin}-\ref{line:derive-permutation-end}).
As presented in Section~\ref{sec:perm_from_seed},
$\Sigma = \op{PermFromRandBits}(G(\var{seed}))$,
where $\var{seed}$ is the XOR of all valid openings for block \bl, 
$G$ is a pseudorandom generator, and $\op{PermFromRandBits}$ an algorithm that derives a permutation from random bits.
The remaining of this block chunks the transactions contained in \bl (L\ref{line:chunk-begin}-\ref{line:chunk-end}), applies $\Sigma$ on the chunked transactions (L\ref{line:permute}), and swaps the first permuted chunk of each of each transaction with the chunk containing the code (L\ref{line:swap}). The function $\op{Chunk}()$ is explained in Section~\ref{sec:chunk}.
Finally, \prot delivers block \bl containing the chunked and permuted transactions through the $\Pi^3\op-{deliver}(\bl)$ event (L\ref{line:deliver-end}).

The function $\MB()$ is an upcall from block-based atomic broadcast. It specifies how a block is filled with transactions and additional data.
For simplicity, the pseudocode omits any detail specific to $\bab$. It first writes all given transactions on the block, then picks uniformly at random $n_\ell$ bit-strings of length $\lambda$.
These are the partial random seeds to be used in the permutation of the following $n_\ell$ blocks, if the block
that is currently being built gets mined and delivered by \bab. The commitments to these partial seeds are appended on the block.

Finally, the predicate $\VB()$ specifies that a block is valid if all its transactions are valid, as specified by $\VT()$,
and if it contains $n_\ell$ commitments. The predicate $\VT()$ is omitted, as its implementation does not affect \prot.

\section{Analysis}\label{sec:analysis}

\subsection{Security analysis}

We model the adversary as an interactive Turing machine (ITM) that corrupts up to $t$ parties at the beginning of the execution. Corrupted parties follow the instructions of the adversary and may diverge arbitrarily from the execution of the protocol. The adversary also has control over the \emph{diffusion functionality}. That is, she can schedule the delivery of messages (within the $\Delta$ rounds), as well as read the $\op{RECEIVE}_i$ of every party at any moment of the execution and directly write in the $\op{RECEIVE}_i$ of any party.

We first show that the security of our construction is derived from the security of the original protocol. Given an execution of protocol $\Pi^3$, we define the equivalent execution in protocol $\Pi$ as the execution in which every party follows the same steps but the commitment, opening, and randomization of transactions are omitted. We also recall the parameters $\tau_1$ and $\tau_2$ that denote the length (in blocks) of the silent and loud phase respectively.

\begin{lemma}
\label{lemma:waiting1}
The probability that an adversary can rewrite a block after any honest partial permutations have been opened is negligible in $\tau_1$. 
\end{lemma}

\begin{proof}
Assume an adversary controlling up to $t$ parties and a block $B$. We know that if $\tau_1>d$, protocol $\Pi$ would deliver block $B$, thus an adversary cannot revert the chain to modify the order of the transactions stored in $B$ but with negligible probability. 
\end{proof}

\begin{lemma}
\label{lemma:waiting2}
 The probability that an adversary can rewrite a chain omitting the opening of some honest partial permutation is negligible in $\tau_2$.
\end{lemma}

\begin{proof}
The fairness quality of protocol $\Pi$ states that for any consecutive $N$ blocks, if $N\geq N_0$ the fraction of honest blocks is at least $\mu$. Thus, if $\tau_2\geq \max\{N_0,\frac{1}{\mu}\}$, there exists at least one honest block containing every opening that is not previously included in the chain. Since 
\end{proof}
 
Our construction aims to turn any protocol into a protocol robust against sandwich attacks. However, there might be new vulnerabilities. Intuitively, our construction should not introduce any vulnerability because the only modified aspect is the order in which transactions are delivered. Theorem~\ref{theo:security} formalizes this intuition.

\begin{remark}
    \label{remark:equivalence}
Note that every $\Pi\op{-delivered}$ block is also $\Pi^3\op{-delivered}$ some block after (Line\ref{line:babdel-begin}--\ref{line:deliver-end}). Note also that every $\Pi^3\op{-delivered}$ is also $\Pi\op{delivered}$. Furthermore, the blocks are delivered in the same order.
\end{remark}

\begin{theorem}
\label{theo:security}
If protocol $\Pi$ implements block-based atomic broadcast, then the Partitioned and Permuted Protocol $\Pi^3$ implements block-based atomic broadcast.
\end{theorem}
\begin{proof}

According to Remark~\ref{remark:equivalence}, the set of $\Pi^3\op{-delivered}$ blocks is the same as the set of $\Pi\op{-delivered}$ blocks.
\begin{description}
  
\item[Validity.] Assume that an honest party $\Pi^3\op{-broadcasts}(\var{tx})$ transaction $\var{tx}$. The party first $\Pi\op{-broadcasts}(\var{tx})$ (L\ref{line:bcast-begin}--\ref{line:bcast-end}). The validity property of protocol $\Pi$ guarantees that eventually a block $\bl$ containing transaction $\var{tx}$ is $\Pi\op{-delivered}$. According to Remark~\ref{remark:equivalence}, the honest party eventually $\Pi^3\op{-delivers}$ a block containing $\var{tx}$ and $\Pi^3$ satisfies the validity property of block-based atomic broadcast.
  
\item[No-duplication.] Note that $\Pi^3$ delivers the same set of blocks as protocol $\Pi$, Remark~\ref{remark:equivalence}. Thus, the no-duplication property of protocol $\Pi^3$ is inherited directly from the no-duplication of protocol $\Pi$.

\item[Agreement.] Consider two honest parties $P$ and $Q$ such that party $P$ $\Pi^3\op{-delivers}$ block $\bl$. Remark~\ref{remark:equivalence} guarantees that $P$ also$\Pi\op{-delivers}$ block $\bl$. The agreement property of protocol $\Pi$ ensure that $Q$ eventually $\Pi\op{-delivers}$ block $\bl$. Remark~\ref{remark:equivalence} guarantees that $Q$ eventually $\Pi^3\op{-delivers}$ block $\bl$. Note that the block $\bl$ delivered by both $P$ and $Q$ may differ in how the transactions are chunked and permuted. However, Lemmas~\ref{lemma:waiting1}~and~\ref{lemma:waiting2} guarantee all correct parties agree on the same permutation with all but negligible probability. Hence, we conclude that protocol $\Pi^3$ satisfies the agreement property.

\item[Total order.] Remark~\ref{remark:equivalence} guarantees that the order in which any honest party $\Pi^3\op{-delivers}$ two block $\bl_1$ and $\bl_2$ is the same as it $\Pi\op{-delivers}$ them. Thus,  the total order property of protocol $\Pi$ guarantees the total order property of protocol $\Pi^3$.
  
\item[External validity.] This follows from the external validity of $\Pi$.
  
\item[Fairness.]According to Remark~\ref{remark:equivalence} the same blocks and in the same order are both $\Pi^3\op{-delivered}$ and $\Pi\op{-delivered}$. Hence, the fairness property of $\Pi^3$ is inherited from the fairness property of protocol~$\Pi$.

\end{description}
\end{proof}

After showing that $\Pi^3$ is as secure as the original protocol $\Pi$. We turn our attention to analyzing the behavior of $\Pi^3$ under sandwich attacks, in the upcoming section.
\subsection{Game-theoretic analysis}
\label{sec:game_analysis}
Here, we aim to show that if we assume all miners are rational, i.e., they prioritize maximizing their own payoff, behaving honestly as according to our protocol $\Pi^3$ is a stable strategy. 

\dotparagraph{Strategic games}
For $N\in\mathbb{N}$, let $\Gamma = (N, (S_i), (u_i))$ be an $N$ party game where $S_i$ is a finite set of strategies for each party $i \in [N]$.
Let $S := S_1 \times \cdots \times S_N$ denote the set of outcomes of the game. 
The utility function of each party $i$, $u_i: S \rightarrow \mathbb{R}$, gives the payoff of party $i$ given an outcome of $\Gamma$.
For any party $i$, a \emph{mixed strategy} $s_i$ is a distribution in $\mu(S_i)$. 
A \emph{strategy profile} of $\Gamma$ is $s := s_1 \times \cdots \times s_N$ where $s_i$ is a mixed strategy of party $i$.
The \emph{expected utility} of a party $i$ given a mixed strategy profile $s$ is defined as $u_i(s) = \mathbb{E}_{a_1 \gets s_1, \cdots, a_N \gets s_N}[u_i(a_1), \cdots, u_i(a_N)]$.
Finally, we note that if $s_i$ is a Dirac distribution over a single strategy $a_i \in S_i$, we say $s_i$ is a \emph{pure strategy} for party $i$.

\dotparagraph{Notation}
Let $w$ denote the total reward for mining a  block and $q$ the negligible probability that a PPT adversary guesses a correct opening.
Recall in~\Cref{sec:reward} that the total block reward $w$ is split between the miner of the block who gets $\alpha \cdot w$ and the miners that append the correct openings who get $\frac{(1-\alpha)\cdot w}{n_\ell}$ for each correct opening they append.
For a given block, we denote by $m$ the number of chunks for \emph{each} transaction in the block, and by $\lambda$ the utility of the sandwich attack on the block.
Specifically, $\lambda$ refers to the utility of a sandwich attack performed on the original transactions in the order they are in \emph{before} chunking and permuting them.
We also denote the optimal sandwich utility by $\Lambda$, which is the maximum utility one can get by performing a sandwich attack.
Finally, we denote by $\hat{\lambda}_i$ the \emph{average} utility of the sandwich attack taken over all blocks on the chain for a specific miner $M_i$. 
This can be computed easily as the transaction mempool is public.
We stress that it is important to look at the average sandwich utility for each miner separately and not the average over all miners as the utility a miner can derive from a sandwich attack depends on their available liquidity (i.e., how much assets they can spare to front-run and back-run the transactions).

\dotparagraph{Quasi-strong $\varepsilon$-Nash Equilibrium}
In terms of game theoretic security, we want our protocols to be resilient to deviations of any subset of miners that form a coalition and deviate jointly.
The security notion we want to achieve is that of an \emph{ quasi-strong $\varepsilon$-Nash Equilibrium}~\cite{Aumann+1959+287+324,BERNHEIM19871,DBLP:conf/icbc2/ChatterjeeGP19}.
Let $C$ denote the coalition of players.
For any strategy profile $s$, we denote by $u_C(s)$ the expected utility of the coalition under $s$.
We denote by $u_C(s'_C, s_{-C})$ the expected utility of the coalition when playing according to some other strategy profile $s'_C$ given the other players that are not part of the coalition play according to $s$.

\begin{definition}(quasi-strong $\varepsilon$-Nash Equilibrium)
\label{def:quasistrongene}
A quasi-strong $\varepsilon$-Nash Equilibrium is a mixed strategy profile $s$ such that for any other strategy profile $s'_C$, $u_C(s) \geq u_C(s'_C, s_{-C}) - \varepsilon$ for some $\varepsilon >0$.
\end{definition}

The notion of a quasi-strong Nash Equilibrium is particularly useful in the context of blockchains as the coalition could potentially be controlled by a single miner with sufficient resources~\cite{DBLP:conf/icbc2/ChatterjeeGP19}. 
The notion of an $\varepsilon$-equilibrium is also important in cases where there could be a small incentive (captured by the $\varepsilon$ parameter) to deviate from the protocol, and of course the smaller one can make $\varepsilon$, the more meaningful the equilibrium.

\dotparagraph{Subgame perfection}
We also consider games that span several rounds and we model them as extensive-form games (see, e.g., \cite{Osborne1994} for a formal definition). 
Extensive form games can be represented as a game tree $\var{tx}$ where the non-leaf vertices of the tree are partitioned to sets corresponding to the players. 
The vertices belonging to each player are further partitioned into information sets $I$ which capture the idea that a player making a move at vertex $x\in I$ is uncertain whether they are making the move from $x$ or some other vertex $x' \in I$.
A subgame of an extensive-form game corresponds to a subtree in $\var{tx}$ rooted at any non-leaf vertex $x$ that belongs to its own information set, i.e., there are no other vertices that are the set except for $x$. 
A strategy profile is a \emph{quasi-strong subgame perfect $\varepsilon$-equilibrium} if it is a quasi-strong $\varepsilon$-Nash equilibrium for all subgames in the extensive-form game. 

\dotparagraph{The induced game}
Let us divide our protocol into epochs: each epoch is designed around a given block say $B_i$ and begins with the generation of random seeds for $B_i$ and ends with appending the openings for the committed random seeds for $B_i$ (i.e., block $B_{i + \tau_1 + \tau_2}$).
We define the underlying game $\Gamma$ induced by any given epoch of our protocol $\Pi^3$.
$\Gamma$ is a $(\tau_2 +1)$-round extensive form game played by $n_\ell + \tau_2$ parties ($n_\ell$ leaders comprising the leader set $L_i$ for any block $B_i$ and the $\tau_2$ miners that mine the blocks $B_{i+\tau_1 + 1} \dots B_{i + \tau_1 + \tau_2}$). 
Note that although we have $N \choose \tau_2$ sets of $\tau_2$ miners to choose from (where $N$ is the total number of miners in the chain) to be the miners of the blocks $B_{i+\tau_1 + 1} \dots B_{i + \tau_1 + \tau_2}$, we can simply fix any set of $\tau_2$ miners together with $L_i$ to be the parties of $\Gamma$ as we assume all miners are rational and so the analysis of the utilities of any set of $\tau_2$ miners will be the same in expectation. 
We use $A$ to denote the set of all miners in $\tau_2$. 
In what follows, we assume an arbitrary but fixed ordering of the miners in $A$.
Round $1$ of $\Gamma$ consists of only the parties in $L_i$ performing actions, namely picking a random seed and committing to it. 
In rounds $2, \dots , \tau_2 +1$ of $\Gamma$, each member of $L_i$ can act by choosing to open their commitment or not.
However, the moment a member of $L_i$ opens its commitment in a given round, they lose the chance to open their commitment in any subsequent round.
Only one miner from $A$ and according to the imposed ordering acts in each round from round $2$ to $\tau_2 +1$ of $\Gamma$. 
The choice of actions of the miner in any of these rounds are the subsets of the set of existing commitment openings (from members of $L_i$) to append to their block.
Finally we note that the $L_i \cap A $ is not necessarily empty and thus miners in the intersection can choose to open and append their commitment in the same round. 

Let us define the honest strategy profile as the profile in which all members of $L_i$ choose to generate a random seed in round $1$ of $\Gamma$, all members of $L_i$ open their commitments at round $\tau_2$ (i.e., at block $B_{i+\tau_1 + \tau_2 -1}$), and each member of $A$ appends all existing opened commitments that appear in the previous round.
We denote the honest strategy profile by $s$.
The security notion we want to achieve for our protocol is a quasi-strong subgame perfect $\varepsilon$-equilibrium (refer to~\Cref{def:quasistrongene}).
Looking ahead, we will also prove that $\varepsilon$ can be made arbitrarily small by increasing the number $m$ of chunks.

\begin{lemma}
\label{lemma:expected}
The expected utility of an honest leader is at least $(1-q)^{n_\ell} \alpha w$.
\end{lemma}
\begin{proof}
The expected utility for a user following the honest strategy comes from the sum of the block reward, the expected utility from the ordering of any of their transactions within the block, and appending valid openings of committed seeds (if any) to their blocks.
The expected utility from the ordering of transactions is $0$ due to symmetry: each possible order is equally likely, for each order that gives some positive utility, there exists a different order producing the same negative utility.
The expected utility from the block reward is $(1-q)^{n_\ell} \alpha w$.
Thus, the total expected utility of an honest miner is at least $(1-q)^{n_\ell}\alpha w$.
\end{proof}

We outline and analyze two broad classes of deviations or attacks any coalition can attempt in this setting. 
The first class happens at round $1$ of $\Gamma$ where the members of the coalition commit to previously agreed seeds to produce a specific permutation of the transactions. 
The coalition then behaves honestly from round $2$ to $\tau_2 +1$ of $\Gamma$. 
We call this attack the \emph{chosen permutation attack} and denote this attack strategy by $s_{CP}$. 
In the second class, the coalition behaves honestly at round $1$ of $\Gamma$, but deviates from round $2$ onwards where some members selectively withhold opening or appending commitments to bias the final permutation.
We call this attack the \emph{biased permutation attack}, and denote it by $s_{BP}$. 

\dotparagraph{Chosen permutation attack}
Before we describe and analyze the chosen permutation attack (for say a block $B_i$), we first show that a necessary condition for the attack to be successful, that is, the coalition's desired permutation happens almost surely, is that at least all $n_\ell$ leaders in $L_i$ have to be involved in the coalition (members of $A$ can also be involved in the coalition, however as we will show this will simply increase the cost).
To do so, we let $S$ denote the set of permutations over the list of transactions and their chunks, and we define what we mean by a protocol $\Pi_{perm}$ (involving $n$ parties) outputs random a permutation in $S$ by the following indistinguishability game called \emph{random permutation indistinguishability} played between a PPT adversary, a challenger, and a protocol $\Pi_{perm}$. First, the adversary corrupts up to $n -1$ parties. The adversary has access to the corrupted parties' transcripts. Then, the challenger samples $\sigma_0$ uniformly at random from $S$, and sets $\sigma_1$ to be the output of $\Pi_{perm}$. After that, the challenger flips a random bit $b$ and sends $\sigma_b$ to the adversary. The game ends with the adversary outputting a bit $b'$. If $b'=b$, the adversary wins the game.
We say a protocol $\Pi_{perm}$ outputs a random permutation if the the adversary wins the above game with probability $\frac{1}{2} + \varepsilon$ for some negligible $\varepsilon$.
Let us define the output of a single round of $\Pi^3$ as the random permutation that is generated from the seeds generated from all leaders in the round according to the algorithm described in~\Cref{sec:perm_from_seed}. 
The following lemma states that as long as a single leader is honest, the output of $\Pi^3$ is pseudorandom.

\begin{lemma}\label{lemma:all}
An adversary that corrupts at most $n_\ell-1$ leaders in a single round of $\Pi^3$ can only win the random permutation indistinguishability game with negligible probability.
\end{lemma}
\begin{proof}
The proof follows in the same way as introduced by M. Blum~\cite{phone}, with the addition of the PRG.
\end{proof}

Lemma~\ref{lemma:all} implies that launching the chosen permutation attack and thus choosing to deviate at round $1$ of $\Gamma$ comes with an implicit cost: either a single miner has to mine $n_\ell$ blocks in a row so the miner single-handedly forms the coalition, or \emph{all} leaders in $L_i$ have to be coordinated into playing according to a predefined strategy. 

\begin{lemma}\label{lemma:single_miner_chosen}
Given the underlying blockchain is secure, the expected utility of the single miner when playing according to $s_{CP}$ is at most $\frac{\lambda}{2^{n_\ell}}$ more than the expected utility of following the honest strategy.
\end{lemma}
\begin{proof}
Since the underlying blockchain is secure, a necessary condition is that a single miner cannot own more than $\frac{1}{2}$ of the total amount of resources owned by all miners of the protocol. 
Thus, the probability of mining $n_\ell$ blocks in a row is strictly less than $\frac{1}{2^{n_\ell}}$.
This means that the expected utility under the attack strategy $u_C(s_{CP}) < \frac{\lambda}{2^{n_\ell}} + n_\ell \alpha w$, which is at most $\frac{\lambda}{2^{n_\ell}}$ larger than the expected utility under the honest strategy which is $u_C(s) = n_\ell \alpha w$.
\end{proof}

The attack strategy of a coalition composed by more than one miner is more complex compared to the case where there is a single miner, as the coalition needs to ensure its members coordinate strategies.
First, the coalition works with the miner of block $B_i$ to select and fix a permutation generated by a specific PRG seed $\sigma_i$. Then, the coalition secret shares $\sigma_i$ among its members\footnote{This not only prevents members from knowing the partial seeds of other members and hence stealing their block reward, but also additionally safeguards the partial seeds of the members against the miner of block $B_i$ who cannot generate a partial seed of their block and hence has nothing to lose.}. After that, the coalition sets up some punishment scheme to penalize members that do not reveal their partial seeds\footnote{This ensures that every member will reveal reveal their partial seeds and the permutation will be generated properly.}. Finally, the coalition commits and reveals these partial seeds in accordance to the protocol $\Pi^3$. 
Let $\mathcal{C}$ denote the expected cost of coordinating the whole chosen permutation attack for the coalition. For this attack to succeed, the expected coordination cost has to be smaller than the expected profit $\lambda$.

\begin{lemma}\label{lemma:mixed_miner_chosen}
The chosen permutation attack fails to be profitable compared to the honest strategy if $\mathcal{C} > \lambda$.
\end{lemma}
\begin{proof}
From Lemma~\ref{lemma:expected}, the expected revenue of an honest miner is $(1-q)^{n_\ell} \alpha w$, thus the expected revenue of the coalition when following the honest strategy is $u_C(s) = n_\ell \cdot (1-q)^{n_\ell} \alpha w$.
The expected revenue for the chosen permutation attack strategy is $u_C(s_{CP}) = n_\ell \cdot (1-q)^{n_\ell} \alpha w + \lambda - \mathcal{C}$. 
Thus, assuming $\mathcal{C} > \lambda$, and since the expected revenue from a mixed strategy is a convex combination of the revenues of the honest and attack strategies, the pure honest strategy gives a strictly larger expected payoff compared to any mixed strategy.
\end{proof}

\begin{remark}
Computing, or even estimating, the coordination cost is non-trivial as it consists of several dimensions and also depends on a myriad of factors and assumptions.
A few notable costs are, firstly, timing costs. 
The coalition has to convince and coordinate all the leaders to agree on a permutation and also commit and reveal them during a short interval of $d$ blocks. 
This involves the cost of securely communicating with all the leaders and also the computational cost involved in setting up the secret sharing scheme.
A second factor is the choice of the initial order of transactions, which the coalition would have to also agree on with the miner of the attacked block. 
Picking transactions greedily would be the simplest choice as finding the optimal set of transactions from the mempool is NP-hard~\cite{DBLP:conf/blockchain2/MeybodiGHS22}. 
Finally, the coalition has to set up a punishment scheme to penalize members that do not reveal their permutations. 
If we ignore the cost of setting up such a scheme, this can be implemented using a deposit scheme with the size of the deposit at least the value of the expected additional per user profit from the sandwich attack~\cite{DBLP:journals/corr/abs-2107-08748}.
This implies an opportunity cost at least linear in $\frac{\lambda}{n_\ell}$, as well as the assumption that each member has at least $\frac{\lambda}{n_\ell}$ to spare to participate in the attack.
Additionally, we note that the coalition could extend to miners from $A$ which are outside the leader set $L_i$. However, since these miners do not contribute to generating the random seeds, they simply add to the communication cost of the coalition.
Finally, we note that the coordination cannot be planned in advance due to the unpredictability of the block mining procedure.
\end{remark}

\dotparagraph{Biased permutation attack}
The intuition behind this attack is that any coalition that controls $k \leq n_\ell$ commitments can choose to select the ones to open or append, which allows the coalition to chose among $2^k$ possible permutations in order to bias the final ordering.
This can be achieved in two situations: either $k$ out of $n_\ell$ leaders of $L_i$ form a coalition and decide which of their commitments to open, or some subset of miners in the loud phase (of size say $\Tilde{k}$) form a coalition and end up controlling $k$ openings, let $\kappa := \min \{k, \Tilde{k}\}$.
Unlike in the case of the chosen permutation attack, it suffices consider the case where we have a \emph{single} miner that happens to either occupy $k$ leader positions among the group of leaders $L_i$ or mine the $\Tilde{k}$ blocks that belong to the coalition in the loud phase.
This is because the case where a coalition of distinct miners that collude only adds additional coordination cost.
The probability that any such coalition gains any additional utility by performing the biased permutation attack compared to the honest strategy can be upper-bounded.
Let $\var{revenue}$ denote the utility the coalition would gain from performing the biased permutation attack.

\begin{lemma}\label{lemma:additional_revenue}
The probability that a coalition of $\kappa$ members performing the biased permutation attack achieves utility of at least $\kappa w >0$ is    
$
\P[\var{revenue}\geq \kappa w]\leq1-(1-e^{-\frac{2m\kappa w}{\lambda}})^{2^k}.
$
\end{lemma}
\begin{proof}
Given a random permutation and a sandwich attack with original utility $\lambda$ (utility if the order of the transactions were not randomized), denote by $\{X_i(\sigma)\}_{i=1}^m$ the utility produced by chunk $i$. The sum of these random variables $X(\sigma)=\sum_{i=1}^n X_i(\sigma)$ represents the total utility of a sandwich attack (after chunking and permuting). $X$ takes values in $[-\lambda,\lambda]$, thus the variables $\{X_i(\sigma)\}$ take values in $[-\frac{\lambda}{m},\frac{\lambda}{m}]$, are equally distributed and are independent. We define the random variables $Y_i(\sigma)=X_i(\sigma)+\frac{\lambda}{m}\in[0,\frac{2\lambda}{m}]$, and $Y(\sigma)=\sum_{i=1}^n Y_i(\sigma)\in[0,2\lambda]$. Using lemma~\ref{lemma:expected}, $\E[Y_i(\sigma)]=\frac{\lambda}{m}$ and
$\E[Y(\sigma)]=\lambda$. Applying Chernoff's bound~\cite{DBLP:books/daglib/0012859} to $Y$,

\begin{equation}
\label{eq:first}
    \P[Y(\sigma)\geq(1+\delta)\E[Y(\sigma)]]\leq e^{\frac{-2\delta^2\E[Y(\sigma)]^2}{m(\frac{\lambda}{m})^2}}=
    e^{-2m\delta^2}
\end{equation}
for $\delta>0$. We can rewrite Equation~\ref{eq:first} as follows:
\begin{equation*}
 \P[\var{revenue}(\sigma)\geq \delta\lambda]=\P[\var{revenue}(\sigma)+\lambda\geq (1+\delta)\lambda]=\P[Y(\sigma)\geq(1+\delta)\E[Y(\sigma)]]\leq e^{-2m\delta^2}.
\end{equation*}
Using the law of total probability we obtain that $\P[\var{revenue}(\sigma)\leq \delta\lambda]\geq 1-e^{-2m\delta^2}.$
Considering the maximum over the $2^k$ possible permutations $\sigma$ and $\delta=\frac{\kappa \omega}{\lambda}$ we conclude that

\begin{equation*}
    \P[\var{revenue}\geq \kappa w]=1-\P[\var{revenue}\leq \kappa w]=1-\P[\var{revenue}(\sigma)\leq \kappa w]\leq 1-(1-e^{-\frac{2m\kappa w}{\lambda}})^{2^k}.
\end{equation*}
\end{proof}

\begin{lemma}\label{lemma:max_prob}
The probability that a coalition of $\kappa$ members has positive additional utility is:
$
\P[\var{revenue}\geq 0]\leq\max_{k'\leq \kappa}\left\{1-(1-e^{-\frac{2mk' w}{\lambda}})^{2^{k}}\right\}.
$
\end{lemma}

\begin{proof}
Lemma~\ref{lemma:additional_revenue} states a bound for the probability that a coalition of $\kappa$ parties has a utility of at least $\kappa w >0$, the penalty for not opening $\kappa$ commitments. Thus, the general case for a coalition aiming to maximize profit is the maximum over $k'\leq \kappa$.
\end{proof}

Recall that $\Lambda$ is the maximal utility and let $p_{k,\lambda}$ denote $\max_{k'\leq \kappa} \{ 1-(1-e^{-\frac{2m(1-q)^{n_\ell}k'w}{\lambda}})^{2^k}\}$.
Then, the expected additional utility from the biased permutation attack of a single miner controlling $k$ leaders is no greater than $p_{k,\lambda}\Lambda$.

Lemmas~\ref{lemma:mixed_miner_chosen},\ref{lemma:additional_revenue}~and~\ref{lemma:max_prob} allow us to prove our main theorem.

\begin{theorem}\label{thm:main}
 Suppose $\mathcal{C} > \lambda$, then the honest strategy $s = ((\text{random seed})_{i=1}^{n_\ell},(\text{open})_{i=1}^{n_\ell})$ is a quasi-strong subgame perfect $\varepsilon$-equilibrium in $\Gamma$ for $\varepsilon = \max\{\frac{\lambda}{2^{n_\ell}}, p_{k,\lambda}\Lambda\}$.   
\end{theorem}
\begin{proof}
We first observe that the expected utility of a coalition that mixes both the chosen and biased permutation attack strategies is no greater than the expected utility of a coalition that performs the chosen permutation attack with a different chosen permutation that accounts for the biasing of the permutation in the second round of $\Gamma$.
Hence, it suffices to analyze the expected utility of the coalition when implementing either of these strategies, i.e., deviating at round $1$ of $\Gamma$ or from rounds $2$ onwards. 

We first analyze the expected utility of a coalition when implementing the chosen permutation attack, which occurs at round $1$ or $\Gamma$.
Since we assume $\mathcal{C} > \lambda$, from Lemma~\ref{lemma:single_miner_chosen} and Lemma~\ref{lemma:mixed_miner_chosen}, we see that any additional expected payoff of any coalition that deviates only at round $1$ of $\Gamma$ by implementing the chosen permutation attack compared to the expected revenue of behaving honestly is at most $\frac{\lambda}{2^{n_\ell}}$.

Now we analyze the expected utility of a coalition when implementing the biased permutation attack. 
From Lemmas~\ref{lemma:additional_revenue}~and~\ref{lemma:max_prob}, we see that the strategy that implements the biased permutation attack across all of rounds $2$ to $\tau_2 +1$ of $\Gamma$ only gives at most $p_{k,\lambda}\Lambda$ more payoff in expectation compared to following the honest strategy $s$ in these rounds.

As such, if we set $\varepsilon = \max\{\frac{\lambda}{2^{n_\ell}}, p_{k,\lambda}\Lambda\}$ to be the largest difference in additional expected revenues between both strategies, we see that $s = ((\text{random seed})_{i=1}^{n_\ell},(\text{open})_{i=1}^{n_\ell})$ is a quasi-strong $\varepsilon$-subgame perfect equilibrium of $\Gamma$.

\end{proof}

\begin{remark}\label{rem:epsilon}
Recall that $\varepsilon$ bounds the additional expected utility an adversary can gain by deviating from the honest strategy profile $s$.
The security of our protocol therefore improves as $\varepsilon = \max\{\frac{\lambda}{2^{n_\ell}}, p_{k,\lambda}\Lambda\}$ decreases.
We observe that the first component $\frac{\lambda}{2^{n_\ell}}$ goes to $0$ exponentially as the size of the leader set $n_\ell$ increases. 
As for the second component $p_{k, \lambda}$, we conduct an empirical analysis of sandwich attacks on Ethereum, \Cref{sec:eth-analysis}, to estimate $p_{k, \lambda}$ and we show that this value approaches zero as the number of chunks $m$ increases.
\end{remark}

\section{Case study: Ethereum MEV attacks}
\label{sec:eth-analysis}
We validate the utility of our results with real-world data from Ethereum.
Specifically, we estimate, using Lemma~\ref{lemma:max_prob}, the probability that a coalition of $k$ parties obtains positive revenue,
for various values of $k$, sandwich revenue $\lambda$, and chunks $m$.
We conclude by analyzing the overhead incurred by protocol \prot as a function of $m$ and its security-efficiency trade-offs. 

\begin{figure}
  \centering
  \includegraphics[width=0.7\textwidth]{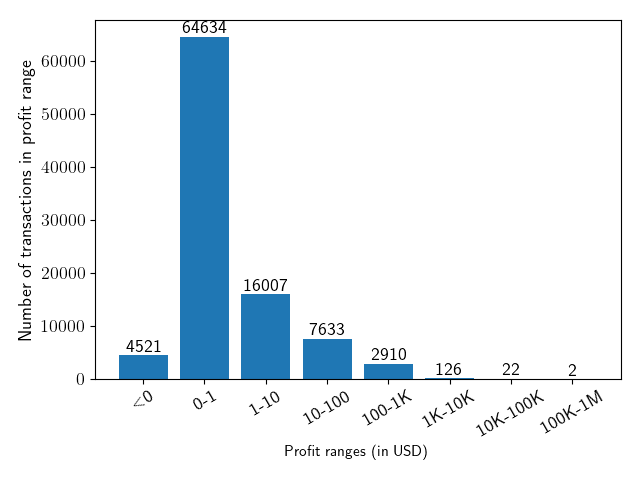}\vspace*{-5mm}
  \caption{
  Detected sandwich attacks on Ethereum in October 2022, grouped by their profit range.
  It can be observed that the majority of the detected attacks had a profit of at most 1 USD,
  and that $99.97\%$ of them had a profit of at most 10K USD.}
  \label{fig:data_october}
\end{figure}

\begin{figure} 
\centering
  \begin{subfigure}{.49\textwidth}
    \centering
    \includegraphics[width=\linewidth]{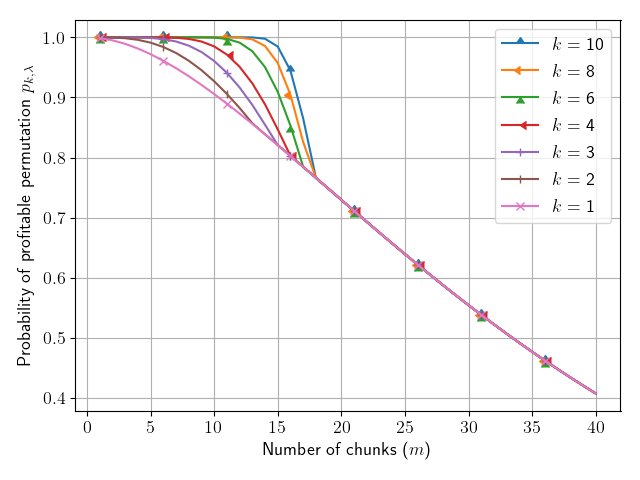}
    \caption{$\lambda = 109 \text{ ETH}$}    
    \label{fig:prob_max}
  \end{subfigure}
  \begin{subfigure}{.49\textwidth}
    \centering
    \includegraphics[width=\linewidth]{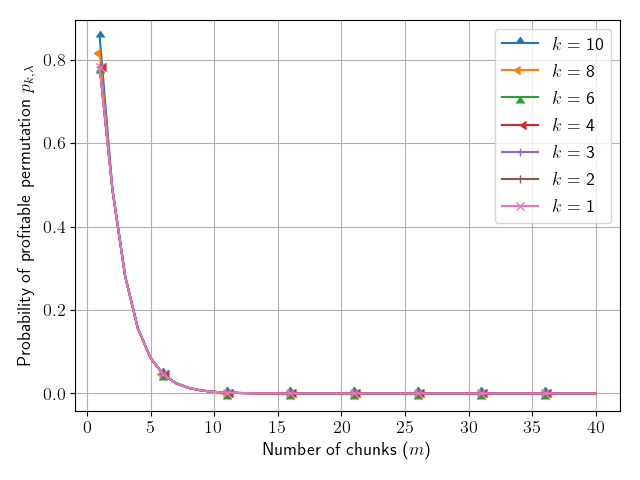}
    \caption{$\lambda = 6.37 \text{ ETH}$}    
    \label{fig:prob_99_97}
  \end{subfigure}%
\caption{The upper bound for the probability of positive revenue $p_{k, \lambda}$ (according to Lemma~\ref{lemma:max_prob})
for a coalition of $k$ leaders, versus the number of chunks $m$.
Fig.~\ref{fig:prob_max} considers $\lambda = 109 \text{ ETH}$, the \emph{maximum} value
among all sandwich attacks detected on Ethereum in October 2022,
and Fig.~\ref{fig:prob_99_97} considers $\lambda = 6.37 \text{ ETH}$, the \emph{$99.97$-th percentile}
of the values.
Fig.~\ref{fig:prob_max} shows that for small number of chunks a larger coalition has a noticeable advantage over a single party,
while for bigger values of $m$ this advantage fades away, as the number of possible permutations grows factorially in $m$, reducing the possible bias of a coalition.
In Fig.~\ref{fig:prob_99_97}, on the other hand, we observe a low probability of positive revenue, even for small values of $m$ (e.g., $ 0.0038$ for $m=10$). Notice that this is the case for 99.97\% of the attacks.
In both figures, parts of the plots coincide. This is because a large $m$ (which implies a factorial increase in the number of possible permutations), combined with the value of the sandwich revenue make it  unprofitable to sacrifice more than one block reward.}
\label{fig:probs}
\end{figure}

\dotparagraph{Empirical security analysis}
We obtain the data on the profit of sandwich attacks on Ethereum using the Eigenphi tool\footnote{\url{https://eigenphi.io/mev/ethereum/sandwich}} for October 2022. We considered October 2022 because it is the month with the highest amount of Sandwich attacks\footnote{https://eigenphi.io/mev/research}, thus its data represents better the state of the art of sandwich attacks. 
To convert between ETH and USD we use the price of ETH as of October 31st, approx. $1,570$ USD.
The block reward at this time is 2 ETH\footnote{\url{https://github.com/ethereum/EIPs/blob/master/EIPS/eip-1234.md}}, or approx. $3,140$ USD.
In Figure~\ref{fig:data_october} we show the number of attacks in bins of increasing profit,
as returned by Eigenphi.
From this data we make use of two facts.
First, $99.97\%$ of the attacks had profit lower than $10$K USD, or approx. $6.37$ ETH,
and second, the most profitable sandwich attack had a profit of 
$170,902.35$ USD, or approx. $109$ ETH.
Hence, we define $\lambda_{99.97} = 6.37$ and $\lambda_{\text{max}} = 109$.
In Figure~\ref{fig:probs} we plot
an upper bound for the probability of positive revenue for a coalition of $k$ leaders,
considering a sandwich revenue of $\lambda_{\text{max}}$ (Figure~\ref{fig:prob_max})
and $\lambda_{99.97}$ (Figure~\ref{fig:prob_99_97}).
We observe that, even for the largest observed sandwich revenue, $\lambda_{\text{max}}$,
the probability of a profitable attack drops below $0.5$ for $m=33$.
For $\lambda_{99.97}$, the probability is low even for small values of $m$. For example, already for
$m = 2$ we get $p_{k, 6.37} \simeq 0.49$, and for $m = 10$ we get $p_{k, 6.37} \simeq 0.0038$, for all $k \geq 1$.
We also remark that Lemma~\ref{lemma:max_prob} states upper bound for the adversary to have \emph{some positive} revenue.

\dotparagraph{Overhead}\label{sec:efficiency_analysis}
In terms of space, 
each block contains exactly $n_\ell$ commitments and on average $n_\ell$ openings of partial seeds.
A commitment to a partial seed takes 256 bits of space. 
Assuming openings are implemented as a call $\op{open}(i,j,\sigma_{i,j})$ to a smart contract,
where $i$ and $j$ are 16-bit integers
and an address is 160 bits long, then each opening consumes 
468 bits on the block.
In total, \prot incurs on average an overhead of 
$724 n_\ell$ bits per block.
As an example, for $n_\ell = 10$, this results in an average overhead of less than 1 KB per block.
We remark that chunking of transactions happens locally,
and hence adds no space overhead on the block.
Concerning execution, there are two main sources of overhead in \prot.
First, when a block is delivered parties compute the final permutation of its $n_t  m$ transactions
using $\op{PermFromRandBits}()$, which has linear-logarithmic bit complexity.
The overhead is thus $O(n_t m \cdot \log(n_t m))$.
Moreover, parties execute $n_t m - n_t $ more transactions,
which incurs an overhead of $O(n_t m)$.
In total, considering $n_t $ to be a constant and $m$ a parameter to \prot,
the execution overhead scales as $O(m \log m)$.
We remark here that the vast majority of computational resources is used in the mining mechanism,
and this is not changed from $\Pi$.
Finally, \prot incurs an increased latency when delivering transactions.
While $\Pi$ has a latency of $d$ blocks, \prot has a latency of $\tau_1 + \tau_2 + d$ blocks.
As discussed earlier, a smaller $\tau_1$ certainly decreases the latency of the protocol, but it also increases the probability of rewriting a block after the commitments that
order its transactions have been opened. Similarly, a smaller $\tau_2$ decreases the latency but gives the miners a shorter time frame to open their commitments. We stress that \emph{not} opening a commitment does not impact the latency of our protocol at all. The only impact it has is on the block reward of the miner who owns the commitment. We also stress that the computational overhead involved in generating the partial seeds in minimum compared with other traditional solutions such as time-lock puzzles~\cite{time-lock}.

\dotparagraph{Security-efficiency tradeoffs}
We first observe a tradeoff between security of \prot and computational overhead.
On the one hand, increasing the number of chunks improves the security of $\Pi^3$.
Recall that $\varepsilon = \max\{\frac{\lambda}{2^{n_\ell}}, p_{k,\lambda}\Lambda\}$, and
in~\Cref{rem:epsilon} we highlight that $p_{k,\lambda}\Lambda$ goes to zero as the number of chunks $m$ increases.
In Figure~\ref{fig:probs} we see how $p_{k,\lambda}$ changes with $m$, based on historical data.
Specifically, for the vast majority of observed sandwich attacks (in Figure~\ref{fig:prob_99_97} we use 
the $99.97$-th percentile) the probability of a coalition to succeed drops exponentially with $m$.
On the other hand, the execution overhead increases as $O(m \log m)$.
Moreover, the size of leader set $n_\ell$ leads to the following tradeoff.
In~\Cref{rem:epsilon} we show that the security of $\Pi^3$ against rational adversaries improves exponentially with $n_\ell$.
However, the size of the leader set also determines the number of leader sets each miner needs to be a part of,
and hence the length of time each miner has to wait until it receives its block reward.

\section{Conclusion}
\label{sec:conclusion}
In this paper we introduced a new construction that can be implemented on top of any blockchain protocol with three main properties. First, the construction does not add any vulnerability to the old protocol, i.e., the security properties remain unchanged. Secondly, performing sandwich attacks in the new protocol is no longer profitable. Thirdly, the construction incurs in minimal overhead with the exception of a minor increase in the latency of the protocol.
Our empirical study of sandwich attacks on the Ethereum blockchain also validates the design principles behind our protocol, demonstrating that our protocol can be easily implemented to mitigate sandwich MEV attacks on the Ethereum blockchain.

\section*{Acknowledgments}

This work has been funded by the Swiss National Science Foundation (SNSF)
under grant agreement Nr\@.~200021\_188443 (Advanced Consensus Protocols).
We would like to thank Krzysztof Pietrzak and Jovana Mićić for useful discussions.

\bibliography{references, dblpbibtex}
\bibliographystyle{ieeesort}

\appendix
\section{Sandwich MEV attacks}\label{app:sandwich_detaildesc}
\dotparagraph{Decentralized exchanges}
Decentralized exchanges (DEXes) allow users to exchange various cryptocurrencies in a decentralized manner (i.e., in a peer-to-peer fashion without a central authority). 
Some examples of DEXes on the Ethereum blockchain are 
Uniswap\footnote{\url{https://uniswap.org}} and Sushiswap\footnote{\url{https://sushi.com}}
DEXes typically function as constant product market makers (CPMMs)~\cite{DBLP:conf/aft/AngerisC20}, i.e., the exchange rate between any two underlying assets is automatically calculated such that the product of the amount of assets in the inventory remains constant.

As an example, consider the scenario where a user at time $t$ wants to swap $\delta_X$ of asset $X$ for asset $Y$ in the $X \rightleftharpoons Y$ liquidity pool, and suppose the pool has $X_t$ and $Y_t$ amount of assets $X$ and $Y$ in its inventory at time $t$.
The user would receive $$\delta_Y = Y_t - \frac{X_T \cdot Y_T}{X_t + (1-f) \delta_X} = \frac{Y_t (1-f) \delta_X}{X_t + (1-f)\delta_X}$$ amount of asset $Y$ for $\delta_X$ amount of asset $X$, where $f$ is a fee charged by the pool~\cite{DBLP:conf/aft/HeimbachW22}. 
We can thus compute the exchange rate of $X$ to $Y$ at time $t$ as

\begin{equation}\label{eq:rate}
    \rho^{XY}_t := \frac{\delta_X}{\delta_Y} = \frac{X_t + (1-f) \delta_X}{Y_t(1-f)}
\end{equation}

\dotparagraph{Sandwich attack}
As transactions in a block are executed sequentially, the exchange rate for a swap transaction could depend on where the transaction is located in the block.
From~\Cref{eq:rate}, we note that the exchange rate from $X$ to $Y$ increases with the size of the trade $\delta_X$.
Thus, if a transaction that swaps $X$ for $Y$ occurs \emph{after} several similar $X$ to $Y$ swap transactions, the exchange rate for this particular transaction would increase. 
Consequently, the user which submitted this transaction would pay more per token of $Y$ as compared to if the transaction occurred \emph{before} the other similar transactions.
In a sandwich attack, the adversary (usually miner) manipulates the order of the transactions within a block such that they can profit from the manipulated exchange rates.
Specifically, the adversary is given the list $\mathcal{T}$ of all transactions that can be included in a block and can make two additional transactions $\var{tx}_1$ and $\var{tx}_2$: transaction $\var{tx}_1$ exchanges some amount (say $\delta_X$) of asset $X$ for asset $Y$, and $\var{tx}_2$ swaps the $Y$ tokens from the output of $\var{tx}_1$ back to $X$.
Let us denote the amount of tokens of $X$ the adversary gets back after $\var{tx}_2$ by $\delta'_X$.
The goal of the adversary is to output a permutation over $\mathcal{T} \cup \{\var{tx}_1, \var{tx}_2\}$ such that $\delta'_X - \delta_X$ is maximized.
A common technique is to front-run all transactions exchanging $X$ to $Y$ in the block, i.e., place transaction $\var{tx}_1$ before all transactions exchanging $X$ to $Y$ and $\var{tx}_2$ after~\cite{DBLP:conf/asiaccs/HeimbachW22}.
We note that users can protect themselves by submitting a slippage bound $\var{sl}>0$ together with each transaction. However, this protection is only partial.
 
\section{Sandwich attacks with random permutation}\label{sec:sandwich_randomperm}
Here we outline a way an adversary can still launch a sandwich attack even when the transactions in a block are randomly permuted by carefully specifying slippage bounds.

\dotparagraph{Background for attack} 
The setting of the attack is as follows: suppose there is a particularly large transaction $t^*$ (that hence impacts the exchange rates) swapping $X$ for $Y$ in the list of transactions $\mathcal{T}$ in a block, and suppose the adversary is aware of $t^*$ (maybe due to colluding with the miner of the block).
We make two further simplifying assumptions: first, that all other transactions are small and hence have negligible impact on the $X \rightleftharpoons Y$ exchange rates, and second, that the swap fee $f$ is negligible.
Like in the case of the classic block sandwich attack, the adversary can create $2$ transactions $\var{tx}_1$ and $\var{tx}_2$ (together with slippage bounds), where $\var{tx}_1$ exchanges some amount of $X$ for $Y$, and $\var{tx}_2$ exchanges the $Y$ tokens back to $X$.
Unlike in the case of the classic sandwich attack, the adversary has no control over the final order of the transactions in the block as the transactions will be randomly permuted. 
An advantageous permutation for the adversary would be any permutation such that $\var{tx}_1$ comes before $t^*$ and $t^*$ comes before $\var{tx}_2$ (hereafter we use the notation $a \prec b$ to denote $a$ ``comes before" $b$ for two transactions $a$ and $b$). 
Permutations that would be disadvantageous to the adversary would be any permutation such that $\var{tx}_2 \prec t^* \prec \var{tx}_1$, as this is the precise setting where the adversary would lose out due to unfavorable exchange rates. 
Any other permutations outside of these are acceptable to the adversary.

\dotparagraph{Utilities} Suppose both $\var{tx}_1$ and $\var{tx}_2$ are executed.
We assume the utility of the adversary is $\alpha \in \mathbb{R}^+$ if the resulting permutation is advantageous, $-\alpha$ if the resulting permutation is disadvantageous, and $0$ for all other permutations.
We assume the utility of the adversary is $0$ if both their trades did not execute, as fees are negligible.
We further assume the following utilities if only $1$ trade executes:
if only $\var{tx}_1$ executes, the utility of the adversary is $0$ if $t^* \prec \var{tx}_1$ and $0 \leq \beta < \alpha$ if $\var{tx}_1 \prec t^*$. 
The intuition behind this is that if $\var{tx}_1$ executes before $t^*$ in this block, there is a chance that when $\var{tx}_2$ executes in the next block the adversary can still benefit from the favorable exchange rates due to advantageous permutation (albeit split over more than $1$ block, thus the discount in utility). 
In the same vein, if only $\var{tx}_2$ executes, the utility of the adversary is $0$ if $t^* \prec \var{tx}_2$ and $ -\gamma < -\alpha$ if $\var{tx}_2 \prec t^*$.
The reason why $\gamma > \alpha$ is to not only take into account the potential loss to the adversary from the disadvantageous permutation, but also the opportunity cost of waiting more than $1$ block for $\var{tx}_1$ to execute.

Here, we denote by $s$ the strategy where the adversary simply wants both transactions to execute and thus does not care about the slippage to be the strategy where the slippage bound for both transactions are set to $\infty$. 
It is clear that the expected utility of the adversary under strategy $s$ is $0$ due to the fact that both advantageous and disadvantageous permutations occur with equal probability. 

\dotparagraph{Sandwich attack by controlling slippage}
We first describe the first attack where the adversary can gain positive expected utility just by being more precise in specifying slippage bounds.
The intuition behind this attack is that by specifying the slippage bounds to be extremely precise, the adversary can ensure that transaction $\var{tx}_1$ always executes before $t^*$, and $\var{tx}_2$ only executes if $\var{tx}_1$ and $t^*$ have executed.

The attack strategy (denoted by $s_{slip}$) is as follows:
\begin{itemize}
    \item The adversary computes the current exchange rate of $X$ to $Y$, denoted by $\rho^{XY}_{\var{tx}_0}$, and sets the slippage bound on $\var{tx}_1$ to be $\rho^{XY}_{\var{tx}_0} + \varepsilon_1$ for some $\varepsilon_1 > 0$. 
    \item The adversary computes the hypothetical exchange rate of $X$ to $Y$ \emph{after} an execution of $t^*$. We denote this rate by $\rho^{XY}_{t^*}$. The adversary also computes the hypothetical exchange rate of $Y$ to $X$ \emph{after} an execution of $\var{tx}_1$, $t^*$, and \emph{both} $\var{tx}_1$ and $t^*$. We denote these rates by $\rho^{XY}_{\var{tx}_1}$, $\rho^{XY}_{t^*}$, and $\rho^{XY}_{\var{tx}_1 + t^*}$ respectively.
    \item The adversary sets the slippage bound for $\var{tx}_2$ to be $\rho^{YX}_{t^*} + \varepsilon_2$ for some other $\varepsilon_2 > 0$.
\end{itemize}

\begin{theorem}
If $\varepsilon_1 < \rho^{XY}_{t^*} - \rho^{XY}_{\var{tx}_0}$ and $\varepsilon_2 < \min(\rho^{YX}_{\var{tx}_1}, \rho^{YX}_{t^*}) - \rho^{YX}_{\var{tx}_1 + t^*}$, the expected utility of $\mu_{slip}$ is $\frac{\alpha}{6} + \frac{\beta}{3} > 0$.
\end{theorem}
\begin{proof}
Let $\sigma$ be a random permutation over $\mathcal{T} \cup \{\var{tx}_1,\var{tx}_2\}$. 
We denote by $[\sigma_{r,i}]$ the position/index of the $i$th transaction after the permutation.

We will proceed case by case for each of the $6$ different orderings of $\var{tx}_1, \var{tx}_2, t^*$.

\begin{itemize}
    \item $\var{tx}_1 \prec t^* \prec \var{tx}_2$: both $\var{tx}_1$ and $\var{tx}_2$ would be executed. The utility of the adversary is $\alpha$ in this case.
    \item $\var{tx}_1 \prec \var{tx}_2 \prec t^*$: $\var{tx}_1$ will be executed. 
    However, $\var{tx}_2$ will not be executed as $\rho^{YX}_{\var{tx}_1 + t^*} + \varepsilon_2 < \rho^{YX}_{\var{tx}_1}$.
    Since $\var{tx}_1 \prec t^*$, the utility of the adversary is $\beta$ in this case. 
    \item $t^* \prec \var{tx}_1 \prec \var{tx}_2$: $\var{tx}_1$ will not be executed as $\rho^{XY}_{\var{tx}_0} + \varepsilon_1 < \rho^{XY}_{t^*}$. 
    $\var{tx}_2$ will be executed. 
    Since $t^* \prec \var{tx}_2$, the utility of the adversary in this case is $0$.
    \item $t^* \prec \var{tx}_2 \prec \var{tx}_1$: both $\var{tx}_1$ and $\var{tx}_2$ will not execute as $\rho^{XY}_{\var{tx}_0} + \varepsilon_1 < \rho^{XY}_{t^*}$ and $\rho^{YX}_{\var{tx}_1 + t^*} + \varepsilon_2 < \rho^{YX}_{t^*}$.
    Since both trades will not execute, the utility of the adversary in this case is $0$.
    \item $\var{tx}_2 \prec \var{tx}_1 \prec t^*$: $\var{tx}_1$ will execute but $\var{tx}_2$ will not execute as $\rho^{YX}_{\var{tx}_1 + t^*} + \varepsilon_2 < \rho^{YX}_{\var{tx}_0}$. 
    Since $\var{tx}_1 \prec t^*$, the utility of the adversary is $\beta$ in this case.
    \item $\var{tx}_2 \prec t^* \prec \var{tx}_1$: both $\var{tx}_1$ and $\var{tx}_2$ will not execute as $\rho^{XY}_{\var{tx}_0} + \varepsilon_1 < \rho^{XY}_{t^*}$ and $\rho^{YX}_{\var{tx}_1 + t^*} + \varepsilon_2 < \rho^{YX}_{\var{tx}_0}$. 
    Since both trades will not execute, the utility of the adversary in this case is $0$.
\end{itemize}

Since each ordering is equally likely to occur, the expected utility of the adversary is $\frac{\alpha}{6} + \frac{\beta}{3} >0$.
\end{proof}

\dotparagraph{Long-range sandwich attacks}
Long-range sandwich attacks are attacks where an adversary aims to front and back run transactions over multiple blocks. 
This can happen when the adversary mines more than $1$ block in a row. 
However, as the probability of mining more than 1 consecutive block is very small (and grows exponentially smaller in the number of consecutive blocks), the success probability of such an approach is similarly low.

An approach that would lead to a larger probability of success would be for the adversary to create two transactions $\var{tx}_1$ and $\var{tx}_2$ and split $\var{tx}_1$ and $\var{tx}_2$ into separate blocks such that $\var{tx}_2$ only conditionally executes upon $\var{tx}_1$ being on the chain. 
Formally, instead of adding both $\var{tx}_1$ and $\var{tx}_2$ to the transaction list $\mathcal{T}$ like in the above attack setting, the adversary now only adds $\var{tx}_1$ to $\mathcal{T}$ and waits until $\var{tx}_1$ is on the chain to add $\var{tx}_2$ to the transaction mempool.
We note that this can be done by wrapping transactions into smart contracts, which can handle the conditional execution of transactions based on some state of the blockchain. 
This can also be done in Bitcoin-like blockchains by ensuring that the UTXO of $\var{tx}_1$ is given as input to $\var{tx}_2$.

The attack strategy (denoted by $s_{longslip}$) is as follows:
\begin{itemize}
    \item The adversary computes the current exchange rate of $X$ to $Y$, denoted by $\rho^{XY}_{\var{tx}_0}$, and sets the slippage bound on $\var{tx}_1$ to be $\rho^{XY}_{\var{tx}_0} + \varepsilon_1$ for some $\varepsilon_1 >0$. 
    \item The adversary waits until $\var{tx}_1$ has been executed (i.e. when the block which contains $\var{tx}_1$ is gossiped), then either wraps $\var{tx}_2$ into a smart contract that checks if $\var{tx}_1$ is on the blockchain and if so executes $\var{tx}_2$, or ensures that the UTXO of $\var{tx}_1$ is given as input to $\var{tx}_2$.
\end{itemize}

Recall that if $\var{tx}_1 \prec t^*$ in a block and $\var{tx}_2$ executes in some block after the block containing $\var{tx}_1$, the utility of the adversary is $0 \leq \beta < \alpha$, and that the utility of the adversary is $0$ if both trades do not occur.
We now show that the expected utility under $s_{longslip}$ is also positive. 

\begin{theorem}\label{thm:longslip}
If $\varepsilon_1 < \rho^{XY}_{t^*} - \rho^{XY}_{\var{tx}_0}$, the expected utility of $s_{longslip}$ is $\frac{\beta}{2}$.
\end{theorem}

\begin{proof}
We note that the orderings $\var{tx}_1 \prec t^*$ and $t^* \prec \var{tx}_1$ are equally likely to occur. 
If $\var{tx}_1 \prec t^*$, $\var{tx}_1$ would be executed together with $t^*$ and thus $\var{tx}_2$ would also be executed in some block after the block containing $\var{tx}_1$.
The expected utility of the adversary is $\beta$ in this case.
If $t^* \prec \var{tx}_1$, $\var{tx}_1$ would not be executed $\rho^{XY}_{\var{tx}_0} + \varepsilon_1 < \rho^{XY}_{t^*}$.
Since $\var{tx}_1$ did not execute, $\var{tx}_2$ would not be executed and thus the utility of the adversary in this case is $0$.
\end{proof}

\begin{remark}
We can make the computation of expected utilities more precise by assuming that the utility of the adversary if $\var{tx}_2$ is executed one block after $\var{tx}_1$ is $\beta$, and multiply $\beta$ by $\delta^{d}$ for some discount factor $\delta <1$ if $\var{tx}_2$ is executed $d$ blocks after $\var{tx}_1$.
However, this would require detailed assumptions about the probability of transactions being selected from the mempool which can depend on fees and other factors. 
This is beyond the scope of our paper, thus we leave this as an interesting direction of future work.
\end{remark}

\end{document}